\documentclass[aps,prd,amsmath,two column,nofootinbib,amssymb,referee]{revtex4}
\usepackage{amssymb}
\usepackage{txfonts}
\usepackage{amsfonts}
\usepackage{mathrsfs}
\usepackage{epsfig,bm,dcolumn}
\usepackage{graphicx}
\usepackage{color}
\usepackage{amsmath}
\usepackage{dcolumn}
\usepackage{overpic}
\usepackage{slashed}

\begin{document}
\title{Kosterlitz-Thouless transition and vortex-antivortex lattice melting in two-dimensional Fermi gases with $p$- or $d$-wave pairing}

\author{Gaoqing Cao$^{1,2}$, Lianyi He$^{3}$ and Xu-Guang Huang$^{2,4}$}
\affiliation{1  School of Physics and Astronomy, Sun Yat-Sen University, Guangzhou 510275, China\\
2 Department of Physics and Center for Particle Physics and Field Theory, Fudan University, Shanghai 200433, China\\
3 State Key Laboratory of Low-Dimensional Quantum Physics and Department of Physics, Tsinghua University, Beijing 100084, China\\
4 Key Laboratory of Nuclear Physics and Ion-beam Application (MOE), Fudan University, Shanghai, China 200433.}

\date{\today}

\begin{abstract}
We present a theoretical study of the finite-temperature Kosterlitz-Thouless (KT) and vortex-antivortex lattice (VAL) melting transitions in two-dimensional Fermi gases with $p$- or $d$-wave pairing. For both pairings, when the interaction is tuned from weak to strong attractions, we observe a quantum phase transition from the Bardeen-Cooper-Schrieffer (BCS) superfluidity to the Bose-Einstein condensation (BEC) of difermions. The KT and VAL transition temperatures increase during this BCS-BEC transition and approach constant values in the deep BEC region. The BCS-BEC transition is  characterized by the non-analyticities of the chemical potential, the superfluid order parameter, and the sound velocities as functions of the interaction strength at both zero and finite temperatures; however, the temperature effect tends to weaken the non-analyticities comparing to the zero temperature case. The effect of mismatched Fermi surfaces on the $d$-wave pairing is also studied.
\end{abstract}


\maketitle
\section{Introduction}

It was proposed by Eagles \cite{Eagles1969} and Leggett \cite{Leggett1980} several decades ago that,  in a
many-fermion system with attractive interaction, one can realize an evolution from the Bardeen-Cooper-Schrieffer (BCS) superfluidity to Bose-Einstein condensation (BEC) of difermion molecules by gradually increasing the strength of the interaction. For $s$-wave interaction, such a BCS- BEC evolution is a smooth crossover \cite{Nozieres1985,SadeMelo1993,Engelbrecht1997,Randeria1989,Randeria1990,Loktev2001,Chen2005,Giorgini2008,Gurarie2007} which has been experimentally studied by using the dilute ultracold fermionic atoms \cite{Greiner2003,Jochim2003,Zwierlein2003}, where the interaction strength is tuned by means of the Feshbach resonance. Such a dilute ultracold atomic system is characterized by a dimensionless parameter $1/(k_{\rm F}a_s)$, where $a_s$ is the $s$-wave scattering length of the short-range interaction and $k_{\rm F}$ is the Fermi momentum in the absence of interaction. The BCS-BEC crossover occurs when $1/(k_{\rm F}a_s)$ goes from $-\infty$ to $\infty$. In addition, the Anderson-Bogoliubov collective mode of fermionic superfluidity at weak attraction evolves smoothly to the Bogoliubov excitation of weakly repulsive Bose condensate at strong attraction \cite{Engelbrecht1997,Gurarie2007,Combescot2006,Belkhir1992,Belkhir1994}.

On the other hand, for nonzero orbital-angular-momentum pairing, such as $p$- or $d$-wave pairing, the BCS-BEC evolution is not smooth but associated with some quantum phase transition\cite{Read2000,Botelho2005a,Botelho2005b,Gurarie2005,Cheng2005,Iskin2006a,Iskin2006b,Cao2013}. Such a quantum phase transition cannot be characterized by a change of symmetry or the associated order parameter. Instead, different quantum phases can be distinguished topologically \cite{Read2000}. Recently, the $p$-wave Feshbach resonance has been realized in three-dimensional ultracold Fermi gases of $^{40}$K~\cite{Luciuk2016} and bosonic $^{85}{\rm Rb}-^{87}$Rb mixture~\cite{Dong2016}, and some of the predicted universal relations for $p$-wave interaction~\cite{Yoshida2015,Yu2015} were successfully verified. On the other hand, the two-dimensional (2D) systems are of particular interest since the topological $p$-wave pairing state exhibits nonabelian statistics \cite{Read2000} and hence is useful for topological computation. In cold-atom experiments, a quasi-2D Fermi gas can be realized by arranging a one-dimensional optical lattice along the axial direction and a weak harmonic trapping potential in the radial plane, such that fermions are strongly confined along the axial direction and form a series of pancake-shaped quasi-2D clouds \cite{2Dexp1,2Dexp2,2Dexp3,2Dexp4,2Dexp5}.

For 2D fermionic systems with generic $p$-or $d$-wave pairing at zero temperature, the thermodynamic quantities and the velocity of the low-energy collective mode can be non-analytic functions of the two-body binding energy at the BCS-BEC quantum phase transition point where the chemical potential vanishes \cite{Botelho2005a,Botelho2005b,Cao2013}. Interestingly, these non-analyticities are determined solely by the infrared behavior of the interaction potential, i.e., independent of the details of the interaction potential as well as the symmetry associated with the order parameter \cite{Cao2013}. However, the temperature in a realistic ultracold atomic gas is always nonzero. Therefore, it is important to study how these non-analyticities are modified when the temperature is nonzero. In addition, it is well known that the thermal superfluid transition in 2D becomes of the Kosterlitz-Thouless (KT) type and vortex-antivortex lattice (VAL) may also exist at low temperature~\cite{BKT,Nelson1979,Young}. It is thus necessary to study the KT and VAL transitions in 2D fermionic systems with $p$-or $d$-wave pairing. The KT and VAL transitions has been comprehensively studied for 2D Fermi gases with $s$-wave pairing \cite{Loktev2001,BKT-S02,BKT-S03,BKT-S04,BKT-S05,BKT-S06,BKT-S07,BKT-S08} and with spin-orbit coupling \cite{BKT-SOC01,BKT-SOC02,BKT-SOC03,BKT-SOC04,BKT-SOC05}. 

In this work, we present a systematical study of the KT and VAL melting transitions in 2D Fermi gases with $p$- or $d$-wave pairing. We find that the non-analyticities are weakened by finite-temperature effect. In particular, we calculate the sound velocity $\upsilon$ as a function of  temperature and interaction strength (the two-body binding energy). For the $p$-wave pairing, $\upsilon$ is a non-monotonous function of the two-body binding energy, while for the $d$-wave pairing, $\upsilon$ decreases monotonously with the binding energy. The effect of mismatched Fermi surfaces is also studied for the $d$-wave pairing. In the BEC regime, we find that the KT and VAL transition temperature both decrease linearly for large chemical potential imbalance, and a superfluid-normal phase transition occurs when the imbalance reaches a critical value.

The paper is arranged as follows. We present the study of $p$-wave pairing system and $d$-wave pairing system in Sec.\ref{chapterp} and Sec.\ref{chapterd}, respectively. The theoretical formalism is given in Sec.\ref{sectionpA} and Sec.\ref{sectiondA}. The numerical results are given in Sec.\ref{sectionpB} for $p$-wave pairing and in Sec.\ref{sectiondB} for $d$-wave pairing. Finally, we summarize in Sec.\ref{summary}. We use the natural units $\hbar=k_B=1$ throughout.

\section{$p$-wave pairing in spinless fermi gases}\label{chapterp}

\subsection{Formalism in Gaussian approximation}\label{sectionpA}

Since the fermion wave function should be anti-symmetric, the simplest setup to study $p$-wave pairing is a ``spinless" Fermi gas, or single-component Fermi gas. The Hamiltonian can be written as~\cite{Read2000,Botelho2005a}
\begin{eqnarray}
{\cal H}&=&\sum_{{\bf k}}\xi_{{\bf k}}^{\phantom{\dag}}\psi_{{\bf k}}^\dagger\psi_{{\bf k}}^{\phantom{\dag}}+\sum_{{\bf k,k',q}}V_{{\bf kk'}}^{\rm p}b_{{\bf kq}}^\dagger b_{{\bf k'q}}^{\phantom{\dag}},
\end{eqnarray}
where $\psi_{{\bf k}}$ represents the fermion annihilation operator, $b_{{\bf kq}}=\psi_{{\bf -k+q}/2}\psi_{{\bf k+q}/2}$, and $\xi_{{\bf k}}=\epsilon_{{\bf k}}-\mu$ with the kinetic energy $\epsilon_{{\bf k}}={\bf k}^2/(2m)$. 
For the sake of simplicity, we consider a separable $p$-wave interaction potential $V_{{\bf kk'}}^{\rm p}$~\cite{Cao2013},
\begin{eqnarray}
V_{{\bf kk'}}^{\rm p}=-\lambda\Gamma^{\rm p}({\bf k})\Gamma^{\rm p*}({\bf k'}),
\end{eqnarray}
where $\lambda$ is the interaction strength.  The gamma functions takes the Nozieres-Schmitt-Rink (NSR) form~\cite{Nozieres1985, Botelho2005a},
\begin{eqnarray}
\Gamma^{\rm p}_s({\bf k})={(k_x+ik_y)/k_1\over(1+k/k_0)^{3/2}},\ \Gamma^{\rm p}_a({\bf k})={k_x/k_1\over(1+k/k_0)^{3/2}},
\end{eqnarray}
with $k=|{\bf k}|$. Here $s$ and $a$ represent the symmetric (isotropic) $p_x+ip_y$ and asymmetric (anisotropic) $p_x$ pairings, respectively. The parameters $k_0$ and $k_1$ set the momentum scale in the short  and long wavelength limits, respectively~\cite{Botelho2005a}. The form of the denominator is chosen to mimic the amplitude damping for $p$-wave partial potential at large momentum~\cite{Botelho2005a}.

The partition function at finite temperature can be given by the imaginary-time path integral formalism, 
\begin{eqnarray}
{\cal Z}=\int [d{\psi^\dagger}][d{\psi}]\exp\left\{-\int_0^\beta d\tau\left(\sum_{{\bf k}}\psi_{{\bf k}}^\dagger\partial_\tau\psi_{{\bf k}}^{\phantom{\dag}}+{\cal H}\right)\right\},
\end{eqnarray}
where $\tau=it$ is the imaginary time and $\beta=1/T$ with $T$ being the temperature.  Introducing an auxiliary bosonic field $\phi_{\bf q}(\tau)=2\lambda\sum_{{\bf k}}\Gamma^{\rm p}({\bf k})b_{{\bf kq}}$ and applying the Hubbard-Stratonovich transformation, we can rewrite the partition function as
\begin{eqnarray}\label{bosonization}
{\cal Z}&=&\int [d{\phi^*}][d{\phi}][d{\Psi^\dagger}][d{\Psi}]\exp\Bigg\{-\int_0^\beta d\tau\bigg[\sum_{\bf q}{|\phi_{\bf q}(\tau)|^2\over4\lambda}\nonumber\\
&&+{1\over2}\sum_{{\bf k,k'}}\Big(\xi_{\bf k}\delta_{\bf k,k'}-\Psi_{{\bf k}}^\dagger G^{-1}_{\bf k,k'}\Psi_{{\bf k'}}^{\phantom{\dag}}\Big)\bigg]\Bigg\},
\end{eqnarray}
where we use the Nambu-Gor'kov representation $\Psi_{{\bf k}}^\dagger=(\psi_{{\bf k}}^\dagger,\psi_{-{\bf k}}^{\phantom{\dag}})$. The inverse fermion Green's function is given by
\begin{eqnarray}
G^{-1}_{\bf k,k'}(\tau)
\!=\!\left(\begin{array}{cc}
\!\!(-\partial_\tau\!-\!\xi_{{\bf k}})\delta_{\bf k,k'}&\!\!\phi_{\bf k-k'}(\tau)\Gamma^{\rm p}\left({{\bf k}+{\bf k'}\over2}\right)\\
\!\!\phi_{\bf -k+k'}^*(\tau)\Gamma^{{\rm p}*}\left({{\bf k}+{\bf k'}\over2}\right)&\!\!(-\partial_\tau\!+\!\xi_{{\bf k}})\delta_{\bf k,k'}
\end{array}\right).
\end{eqnarray}
Integrating out the fermion degrees of freedom, we obtain
\begin{eqnarray}
{\cal Z}=\int [d{\phi^*}][d{\phi}]~e^{-{\cal S}_{\rm eff}^{\rm p}[\phi^*,\phi]},
\end{eqnarray}
with the effective action
\begin{eqnarray}
{\cal S}_{\rm eff}^{\rm p}=\int_0^\beta d\tau \ \Bigg[\sum_{\bf q}{|\phi_{\bf q}(\tau)|^2\over4\lambda}+{1\over2}\sum_{{\bf k,k'}}\Big(\xi_{\bf k}\delta_{\bf k,k'}-\text{Tr}\ln G^{-1}_{\bf k,k'}\Big)\Bigg],
\end{eqnarray}
where the trace is taken over imaginary time, momentum and Nambu-Gor'kov spaces.

To proceed, we decompose the auxiliary field $\phi_{\bf q}(\tau)$ into its mean-field and fluctuation parts,
\begin{eqnarray}
\phi_{\bf q}(\tau)=\Delta\delta_{\bf q,0}+\hat\phi_{\bf q}(\tau).
\end{eqnarray}
The effective action can be evaluated in powers of the fluctuation $\hat\phi_{\bf q}(\tau)$, i.e., ${\cal S}_{\rm eff}^{\rm p}={\cal S}_0^{\rm p}+{\cal S}_2^{\rm p}+\cdots$. Here we omit the linear term in the fluctuation since it vanishes due to the gap equation. The leading-order term ${\cal S}_0^{\rm p}$ represents the mean-field contribution. The next-to-leading-order term ${\cal S}_2^{\rm p}$, which is quadratic in the fluctuation, represents the Gaussian fluctuations and hence the collective mode dynamics.

\subsubsection{Mean-field approximation}\label{sectionp1}
The mean-field contribution $S_0^{\rm p}$ can be evaluated as
\begin{eqnarray}
{\cal S}_0^{\rm p}&=&\beta S\left[{\Delta^2\over4\lambda}+{1\over2}\int\!\!\! {d^2{\bf k}\over(2\pi)^2}\xi_{\bf k}\!-\!{T\over2}\!\sum_{n}\!\int\!\!\! {d^2{\bf k}\over(2\pi)^2}\ln \text{det}{\cal G}^{-1}_{\bf k}(i\omega_n)\right]\nonumber\\
&=&{\beta S}\left\{{\Delta^2\over4\lambda}\!-\!\frac{1}{2}\int\!\!\!{d^2{\bf k}\over(2\pi)^2}\left[E_{\bf k}\!-\!\xi_{\bf k}\!+\!2T\ln\big(1\!+\!e^{-E_{\bf k}/T}\big)\right]\right\},
\end{eqnarray}
where $S$ is the area of the system and  $\omega_n=(2n+1)\pi T \ (n\in\mathbb{Z})$ is the fermion Matsubara frequency. The inverse fermion Green's function  in mean-field approximation is given by 
\begin{eqnarray}
{\cal G}^{-1}_{\bf k}(i\omega_n)=\left(\begin{array}{cc}
i\omega_n-\xi_{{\bf k}}&\Delta_{\bf k}^{\rm p}\\
\Delta_{\bf k}^{{\rm p}*}&i\omega_n+\xi_{{\bf k}}
\end{array}\right),
\end{eqnarray}
which gives the fermionic quasiparticle spectrum $E_{\bf k}=\left(\xi_{{\bf k}}^{2}+|\Delta_{\bf k}^{\rm p}|^2\right)^{1/2}$ with $\Delta_{\bf k}^{\rm p}=\Delta\Gamma^{\rm p}({\bf k})$.  The mean field $\Delta$, normally referred to as the superfluid order parameter, is determined by the extreme condition $\partial {\cal S}_0^{\rm p}/\partial\Delta=0$, which gives rise to the gap equation
\begin{eqnarray}\label{gapp}
{1\over\lambda}=\int {d^2{\bf k}\over(2\pi)^2}{|\Gamma^{\rm p}({\bf k})|^2\over E_{\bf k}}\tanh\left({E_{\bf k}\over2T}\right).
\end{eqnarray}
The mean-field contribution to the number density is obtained through the thermodynamic relation $n_0=-(\partial {\cal S}_0^{\rm p}/\partial\mu)/(\beta S)$. We have
\begin{eqnarray}\label{number0}
n_0\equiv\int{d^2{\bf k}\over(2\pi)^2}n_0({\bf k})={1\over2}\int{d^2{\bf k}\over(2\pi)^2}\left[1-{\xi_{{\bf k}}\over E_{\bf k}}\tanh\left({E_{\bf k}\over2T}\right)\right].
\end{eqnarray}

The interaction strength $\lambda$ can be physically characterized by the two-body binding energy $E_b$ in vacuum. It is given by~\cite{Botelho2005a}
\begin{eqnarray}
{1\over\lambda}=\int{d^2{\bf k}\over(2\pi)^2}{2|\Gamma^{\rm p}({\bf k})|^2\over2\epsilon_{{\bf k}}-E_b}.
\end{eqnarray}
Note that unlike the $s$-wave case, here the binding energy $E_b$ can be both negative or positive. The weak and strong attraction limits correspond to
$E_b\rightarrow+\infty$ and $E_b\rightarrow-\infty$, respectively.

\subsubsection{Gaussian fluctuation and Goldstone mode}\label{sectionp2}
The Gaussian fluctuation contribution to the effective action is quadratic in $\phi_{\bf q}(\tau)$ and thus represents the collective mode dynamics. It can be evaluated as
\begin{eqnarray}
{\cal S}_2^{\rm p}&=&\sum_{{\bf q},n}\Bigg\{{|\hat\phi_{\bf q}(i\nu_n)|^2\over4\lambda}+{T\over4S}\!\sum_{{\bf k},m}{\rm Tr}~\bigg[{\cal G}_{{\bf k}-{\bf q}/2}(i\omega_m) \nonumber\\
&&\times~\Phi_{\bf -q}(-i\nu_n){\cal G}_{{\bf k}+{\bf q}/2}(i\omega_m+i\nu_n)\Phi_{\bf q}(i\nu_n)\bigg]\Bigg\},
\end{eqnarray}
where $\nu_n=2\pi nT ~(n\in\mathbb{Z})$ is the boson Matsubara frequency, and the mean-field fermion Green's function and the vertex matrix $\Phi$ are given by
\begin{eqnarray}
{\cal G}_{\bf k}(i\omega_m)&=&{1\over(i\omega_m)^2-E_{{\bf k}}^2}\left(\begin{array}{cc}
(i\omega_m+\xi_{{\bf k}})&-\Delta_{\bf k}^{\rm p}\\
-\Delta_{\bf k}^{{\rm p}*}&(i\omega_m-\xi_{{\bf k}})
\end{array}\right),\nonumber\\
\Phi_{\bf q}(i\nu_n)&=&\left(\begin{array}{cc}
0&\!\!\hat\phi_{\bf q}(i\nu_n)\Gamma^{\rm p}({\bf k})\\
\hat\phi_{\bf -q}^*(-i\nu_n)\Gamma^{\rm p*}({\bf k})&0
\end{array}\right).
\end{eqnarray}
After some algebra, ${\cal S}_2^{\rm p}$ can be written in a compact form 
\begin{eqnarray}
{\cal S}_2^{\rm p}=\frac{1}{2}\sum_{{\bf q},n}\left(\begin{array}{cc}
\hat{\phi}_{\bf q}^*(i\nu_n) & \hat{\phi}_{-{\bf q}}(-i\nu_n)\end{array}\right) M({\bf q},i\nu_n)\left(\begin{array}{cc} \hat{\phi}_{\bf q}(i\nu_n)\\
\hat{\phi}_{-{\bf q}}^*(-i\nu_n)\end{array}\right),
\end{eqnarray}
where the inverse boson propagator $M({\bf q},i\nu_n)$ takes the form
\begin{eqnarray}
M({\bf q},i\nu_n)=\left(\begin{array}{cc}M_{11}({\bf q},i\nu_n)&M_{12}({\bf q},i\nu_n)\\
M_{21}({\bf q},i\nu_n)& M_{22}({\bf q},i\nu_n)\end{array}\right).
\end{eqnarray}
The matrix elements are given by
\begin{eqnarray}
M_{11}&=&{1\over4\lambda}+{T\over2S}\sum_{{\bf k},m}{\cal G}_{{\bf k}-{\bf q}/2}^{11}(i\omega_m){\cal G}_{{\bf k}+{\bf q}/2}^{22}(i\omega_m+i\nu_n)|\Gamma^{\rm p}({\bf k})|^2,\nonumber\\
M_{22}&=&{1\over4\lambda}+{T\over2S}\sum_{{\bf k},m}{\cal G}_{{\bf k}-{\bf q}/2}^{22}(i\omega_m){\cal G}_{{\bf k}+{\bf q}/2}^{11}(i\omega_m+i\nu_n)|\Gamma^{\rm p}({\bf k})|^2,\nonumber\\
M_{12}&=&{T\over2S}\sum_{{\bf k},m}{\cal G}_{{\bf k}-{\bf q}/2}^{12}(i\omega_m){\cal G}_{{\bf k}+{\bf q}/2}^{12}(i\omega_m+i\nu_n)[\Gamma^{{\rm p}*}({\bf k})]^2,\nonumber\\
M_{21}&=&{T\over2S}\sum_{{\bf k},m}{\cal G}_{{\bf k}-{\bf q}/2}^{21}(i\omega_m){\cal G}_{{\bf k}+{\bf q}/2}^{21}(i\omega_m+i\nu_n)[\Gamma^{{\rm p}}({\bf k})]^2.
\end{eqnarray}
It is easy to prove that these matrix elements satisfy 
\begin{eqnarray}
M_{11}^*({\bf q},i\nu_n)=M_{22}({\bf q},i\nu_n),\ \ M_{12}^*({\bf q},i\nu_n)=M_{21}({\bf q},i\nu_n).
\end{eqnarray}
Completing the fermion Matsubara frequency summation, we obtain 
\begin{widetext}
\begin{eqnarray}\label{matrixp}
M_{11}&=&{1\over4\lambda}+\int\!\!\! {d^2{\bf k}\over(2\pi)^2}{|\Gamma^{\rm p}({\bf k})|^2\over2}\left[\Bigg({u_-^2\upsilon_+^2\over i\nu_n\!+\!(E_+\!-\!E_-)}\!-\!{u_+^2\upsilon_-^2\over i\nu_n-(E_+-E_-)}\Bigg)(f_+\!-\!f_-)\!+\!\Bigg({u_+^2u_-^2\over i\nu_n\!-\!(E_+\!+\!E_-)}\!-\!{\upsilon_+^2\upsilon_-^2\over i\nu_n+(E_+\!+\!E_-)}\Bigg)(1\!-\!f_+\!-\!f_-)\right],\nonumber\\
M_{12}&=&-\int\!\!\! {d^2{\bf k}\over(2\pi)^2}{[\Gamma^{{\rm p}*}({\bf k})]^2\over8E_+E_-}\left[\Bigg({\Delta_{{\bf k}-{\bf q}/2}^{\rm p}\Delta_{{\bf k}+{\bf q}/2}^{\rm p}\over i\nu_n\!-\!(E_+\!-\!E_-)}\!-\!{\Delta_{{\bf k}-{\bf q}/2}^{\rm p}\Delta_{{\bf k}+{\bf q}/2}^{\rm p}\over i\nu_n+(E_+-E_-)}\Bigg)(f_+\!-\!f_-)\!+\!\Bigg({\Delta_{{\bf k}-{\bf q}/2}^{\rm p}\Delta_{{\bf k}+{\bf q}/2}^{\rm p}\over i\nu_n\!-\!(E_+\!+\!E_-)}\!-\!{\Delta_{{\bf k}-{\bf q}/2}^{\rm p}\Delta_{{\bf k}+{\bf q}/2}^{\rm p}\over i\nu_n+(E_+\!+\!E_-)}\Bigg)(1\!-\!f_+\!-\!f_-)\right],
\end{eqnarray}
\end{widetext}
where the BCS distributions are defined as $u_\pm^2=(1+\xi_\pm/E_\pm)/2$ and $\upsilon_\pm^2=(1-\xi_\pm/E_\pm)/2$, and the Fermi-Dirac distribution is given by $f_\pm=\Big(1+e^{E_\pm/T}\Big)^{-1}$, 
with the dispersions $\xi_\pm=\xi_{{\bf k}\pm{\bf q}/2}$ and $E_\pm=E_{{\bf k}\pm{\bf q}/2}$.  We note that the terms proportional to $f_+-f_-$ corresponds to the Landau damping effect, which vanish when $T\rightarrow0$.

It is more physical to decompose the fluctuation into its real and imaginary parts, i.e., $\hat\phi(x)=\sigma(x)+i\pi(x)$. In momentum space we have
$\hat\phi_{\bf q}(i\nu_n) =\sigma_{\bf q}(i\nu_n)+i\pi_{\bf q}(i\nu_n)$ and $\hat\phi_{\bf q}^*(i\nu_n) =\sigma_{\bf q}^*(i\nu_n)-i\pi_{\bf q}^*(i\nu_n)=\sigma_{\bf -q}(-i\nu_n)-i\pi_{\bf -q}(-i\nu_n)$. Thus, the Gaussian fluctuation part of the effective action can be expressed as
\begin{eqnarray}
{\cal S}_2^{\rm p}={1\over2}\sum_{{\bf q},n}\left(\begin{array}{cc}\sigma_{\bf q}^*(i\nu_n)&\pi_{\bf q}^*(i\nu_n)\end{array}\right)\Pi({\bf q},i\nu_n)\left(\begin{array}{c}\sigma_{\bf q}(i\nu_n)\\ \pi_{\bf q}(i\nu_n)\end{array}\right),
\end{eqnarray}
where the inverse boson propagator reads
\begin{eqnarray}
\Pi\!=\!\left(\begin{array}{cc}
\!\!(M_{11}\!+\!M_{12}\!+\!M_{21}\!+\!M_{22})\!&\!\!i(\!-\!M_{11}\!-\!M_{12}\!+\!M_{21}\!+\!M_{22})\!\!\!\\
\!\!i(M_{11}\!-\!M_{12}\!+\!M_{21}\!-\!M_{22})\!&\!\!(M_{11}\!-\!M_{12}\!-\!M_{21}\!+\!M_{22})\!\!\!
\end{array}\right).
\end{eqnarray}
The low-energy dynamics is governed by the gapless Goldstone mode. Diagonalizing the matrix $\Pi$, we obtain two eigen-modes. Their inverse propagators are given by
\begin{eqnarray}
{\cal D}_{\theta/\eta}^{-1}({\bf q},i\nu_n)=M_{11}+M_{22}\mp\sqrt{(M_{11}-M_{22})^2+4M_{12}M_{21}}.
\end{eqnarray}
We can prove that ${\cal D}_{\theta}^{-1}({\bf 0},0)=0$, which indicates that the $\theta$-mode is  gapless, i.e., the Goldstone mode. It is a mixture of $\sigma$ and $\pi$ components and can be expressed as
\begin{eqnarray}
\theta_{\bf q}(i\nu_n)={\cal C}\left[{\cal D}_{\theta}^{-1}({\bf q},i\nu_n)\sigma_{\bf q}(i\nu_n)+{\cal D}_{\eta}^{-1}({\bf q},i\nu_n)\pi_{\bf q}(i\nu_n)\right],
\end{eqnarray}
where ${\cal C}$ is a normalization coefficient.

The KT transition is related to the stiffness of the Goldstone mode, i.e., the gapless $\theta$ mode. To this end, we need to study the low-energy dynamics of the collective modes. At small energy and momentum, 
the propagator of the gapless $\theta$ mode can be expressed as
\begin{eqnarray}\label{GM-EXP}
{\cal D}_\theta^{-1}({\bf q},i\nu_n)=-\zeta^{\rm p}(i\nu_n)^2+\frac{1}{4m\Delta^2}\left(\rho_x^{\rm p}q_x^2+\rho_y^{\rm p}q_y^2\right),
\end{eqnarray}
where $\rho_x^{\rm p}$ and $\rho_y^{\rm p}$ are the so-called stiffnesses of the Goldstone mode.
To compute the coefficients $\zeta^{\rm p}$, $\rho_x^{\rm p}$, and $\rho_y^{\rm p}$, we make the low-energy expansion of $M_{\rm ij}$ (${\rm i,j}=1,2$) to the quadratic order in frequency and momentum, 
\begin{eqnarray}
M_{\rm ij}({\bf q},i\nu_n)=A_{\rm ij}+i\nu_nB_{\rm ij}+(i\nu_n)^2C_{\rm ij}+D_{\rm ij}^xq_x^2+D_{\rm ij}^yq_y^2.
\end{eqnarray}
However, because of the Landau damping terms proportional to $f_+-f_-$ in Eq. (\ref{matrixp}), such an expansion is in principle only valid at zero temperature or near the superfluid transition temperature 
\cite{Engelbrecht1997, EXPANSION}.  Mathematically, the Landau damping terms bring divergences when doing such an expansion. Since the KT and VAL melting transitions occur at low temperature where the pairing gap is still large,  we may neglect the divergences from Landau damping effect and perform this expansion. Physically, in this approximation, we neglect the damping of the collective modes and treat them as stable modes. Below the KT transition temperature, we expect that the large pairing gap suppresses the damping of the collective modes and validates this approximation. 

By neglecting the Landau damping effect, we can evaluate the expansion coefficients as
\begin{eqnarray}\label{ABCp}
&&A_{11}=A_{22}={1\over4\lambda}-\int\!\!\! {d^2{\bf k}\over(2\pi)^2}\left(E_{\bf k}^2+\xi_{\bf k}^2\right){|\Gamma^{\rm p}({\bf k})|^2\over8E_{\bf k}^3}\tanh\left({E_{\bf k}\over2T}\right),\nonumber\\
&&B_{11}=-B_{22}=-\int\!\!\! {d^2{\bf k}\over(2\pi)^2}{\xi_{\bf k}|\Gamma^{\rm p}({\bf k})|^2\over8E_{\bf k}^3}\tanh\left({E_{\bf k}\over2T}\right),\nonumber\\
&&C_{11}=C_{22}=-\int\!\!\! {d^2{\bf k}\over(2\pi)^2}\left(E_{\bf k}^2+\xi_{\bf k}^2\right){|\Gamma^{\rm p}({\bf k})|^2\over32E_{\bf k}^5}\tanh\left({E_{\bf k}\over2T}\right),\nonumber\\
&&A_{12}=A_{21}=\Delta^2\int\!\!\! {d^2{\bf k}\over(2\pi)^2}{|\Gamma^{\rm p}({\bf k})|^4\over8E_{\bf k}^3}\tanh\left({E_{\bf k}\over2T}\right),\nonumber\\
&&B_{12}=B_{21}=0,\nonumber\\
&&C_{12}=C_{21}=\Delta^2\int\!\!\! {d^2{\bf k}\over(2\pi)^2}{|\Gamma^{\rm p}({\bf k})|^4\over32E_{\bf k}^5}\tanh\left({E_{\bf k}\over2T}\right).
\end{eqnarray}
The coefficients $D_{\rm ij}^{x,y}$ can be obtained but quite lengthy (see Appendix. \ref{rho}). Here we show the combined quantities
\begin{eqnarray}
\rho_i^{\rm p}=4m\Delta^2(D_{11}^i+D_{22}^i-D_{12}^i-D_{21}^i), \ (i=x,y)
\end{eqnarray}
which are exactly the superfluid density along the $x$ and $y$ directions. After a lengthy calculation we obtain
\begin{eqnarray}\label{rhop}
\rho_i^{\rm p}=\int\!\!\! {d^2{\bf k}\over(2\pi)^2}\left[n_0({\bf k})-\frac{k_i^2}{4mT}{\rm sech}^2\left({E_{\bf k}\over2T}\right)\right].
\end{eqnarray}
At zero temperature, the superfluid density is isotropic for both $p_x$ and $p_x+ip_y$ pairings and we have $\rho_x^{\rm p}=\rho_y^{\rm p}=n$ as required by the Galilean invariance~\cite{Leggett2006}. However, for $p_x$ pairing, finite temperature effect generates anisotropy of the superfluid density.  

Finally, the low-energy behavior of the $\theta$-mode or the Goldstone mode is given by Eq. (\ref{GM-EXP}), where the coefficient $\zeta^{\rm p}$ reads
\begin{eqnarray}
\zeta^{\rm p}=\zeta_0^{\rm p}+\frac{B_{11}^2}{A_{12}},
\end{eqnarray}
with $\zeta_0^{\rm p}=-2(C_{11}-C_{12})$.  The Goldstone mode velocity or sound velocity along the $i$-direction reads 
\begin{eqnarray}
\upsilon_i^{\rm p}=\sqrt{\frac{\rho_i^{\rm p}}{4m\Delta^2\zeta^{\rm p}}}.
\end{eqnarray}
We note that the term $B_{11}^2/A_{12}$ arises from the coupling between the phase and amplitude modes and is rather important to recover the correct sound velocity in the
BCS-BEC evolution \cite{BKT-S08}. We also emphasize that even though the  low-energy expansion of the matrix elements $M_{\rm ij}({\bf q},i\nu_n)$ suffers from the divergence problem 
caused by the Landau damping effect, these divergences cancels exactly for the coefficients $\zeta^{\rm p}$, $\rho_x^{\rm p}$, and $\rho_y^{\rm p}$. The divergences only arises for higher-order terms in the 
expansion (\ref{GM-EXP}).  These divergences correspond to the damping of the Goldstone mode and we may neglect it at low temperature.

Comparing to previous approach to KT transition in superfluid 2D Fermi gases \cite{Loktev2001,Botelho2006,BKT-S04}, we make some comments here. Previous approach adopted an alternative decomposition of the superfluid order parameter field $\phi(x)$, i.e., $\phi(x)=[\Delta+\eta(x)]e^{i\theta(x)}$, and the amplitude fluctuation $\eta(x)$ is normally neglected \cite{Loktev2001,Botelho2006,BKT-S04}. The KT transition can be obtained by studying the low energy dynamics of the pure phase mode $\theta(x)$. The advantage of this approach is that it formally does not suffer from the Landau damping problem as we encounter here. We have also evaluated the low-energy expansion for the phase mode in this approach. The expansion also takes the form (\ref{GM-EXP}) and leads to the same result for the superfluid density $\rho_i^{\rm p}$.
However,  the coefficient ${\zeta}^{\rm p}$ is different \cite{Botelho2006}:
\begin{eqnarray}
{\zeta}^{\rm p}\!=\!\!\int\!\!\! {d^2{\bf k}\over(2\pi)^2}{|\Gamma^{\rm p}({\bf k})|^2\over8E_{\bf k}^2}\left[{|\Delta_{\bf k}^{\rm p}|^2\over E_{\bf k}}\tanh\left({E_{\bf k}\over2T}\right)\!+\!\frac{\xi_{\bf k}^2}{2T}{\rm sech}^2\left({E_{\bf k}\over2T}\right)\right].
\end{eqnarray}
We can easily identify that the first term is just $\zeta_0^{\rm p}$ and the second term comes from the fact that this approach amounts to take the limit ${\bf q}\rightarrow0$ first when evaluating the low-energy expansion. As clarified in \cite{BKT-S08}, this approach leads to incorrect result for the sound velocity $\upsilon_i^{\rm p}$ in the BCS-BEC evolution.  In summary,  our approach can recover not only the correct superfluid density but also the correct sound velocity. The price we pay in this approach is that we have to neglect the damping of the collective modes.

In our low-energy approximation, the contribution of the Goldstone mode to the thermodynamic potential can be given by
\begin{eqnarray}
\Omega_2^{\rm p}=\int{d^2{\bf q}\over(2\pi)^2}T\ln\Big(1-e^{-\varepsilon_{\bf q}/T}\Big)=-{\zeta(3)T^3\over2\pi(\upsilon_x^{\rm p}\upsilon_y^{\rm p})},
\end{eqnarray}
where the dispersion relation of the Goldstone mode is given by $\varepsilon_{\bf q}=[\sum_{i=x,y}(\upsilon_i^{\rm p}q_i)^2]^{1/2}$ and $\zeta(x)$ is the Riemann zeta function. At finite temperature, we take into account 
the fluctuation contribution to the  number density. The total fermion number density $n$ can be given by
\begin{eqnarray}\label{number}
n=n_0-\frac{\partial \Omega_2^{\rm p}}{\partial\mu}=n_0-{\zeta(3)T^3\over2\pi(\upsilon_x^{\rm p}\upsilon_y^{\rm p})^2}{\partial(\upsilon_x^{\rm p}\upsilon_y^{\rm p})\over\partial\mu}.
\end{eqnarray}
At $T=0$ we have $n=n_0$ and therefore the quantum fluctuations~\cite{GPF01,GPF02,GPF03} are not taken into account in the present theory. For $s$-wave pairing, it was found that inclusion of quantum fluctuations
leads to slight correction to the KT transition~\cite{GPF04,GPF05}. Thus we expect that the present theory can provides reliable results for the KT and VAL transition for higher partial wave pairings.

\subsection{Kosterlitz-Thouless and vortex-antivortex lattice melting transitions}\label{sectionpB}
The KT and VAL melting temperatures are both directly related to the stiffness $J_{\rm i}=\rho_{\rm i}^{\rm p}/(4m)$~\cite{BKT,Nelson1979,Young,Botelho2006}:
\begin{eqnarray}\label{TC}
T_{\rm KT}={\pi\over2}\sqrt{J_{x}^{\rm p}(T_{\rm KT})J_{y}^{\rm p}(T_{\rm KT})},\ \ T_{\rm M}=0.3\sqrt{J_{x}^{\rm p}(T_{\rm M})J_{y}^{\rm p}(T_{\rm M})}.
\end{eqnarray}
For the anisotropic $p_x$ pairing, the vortex might be elliptically shaped and the usual square vortex-antivortex lattice will also deform accordingly just like the case with anisotropic spin-orbit coupling~\cite{BKT-SOC02}. However, one can scale one direction so that the scaled vortex is circular (the scaled lattice thus becomes square). Thus we can apply Eq.~(\ref{TC}) to the scaled vortex and lattice. Then for a given $E_b$ and number density, the gap equation (\ref{gapp}), the number equation (\ref{number}), and the critical temperature equation Eq.(\ref{TC}) can be solved self-consistently to give $T_{\rm KT}$ ($T_{\rm M}$) and $\Delta$ and $\mu$ at $T_{\rm KT}$ ($T_{\rm M}$).

\begin{figure}[!htb]
\centering
\begin{overpic}
[scale=1.4]{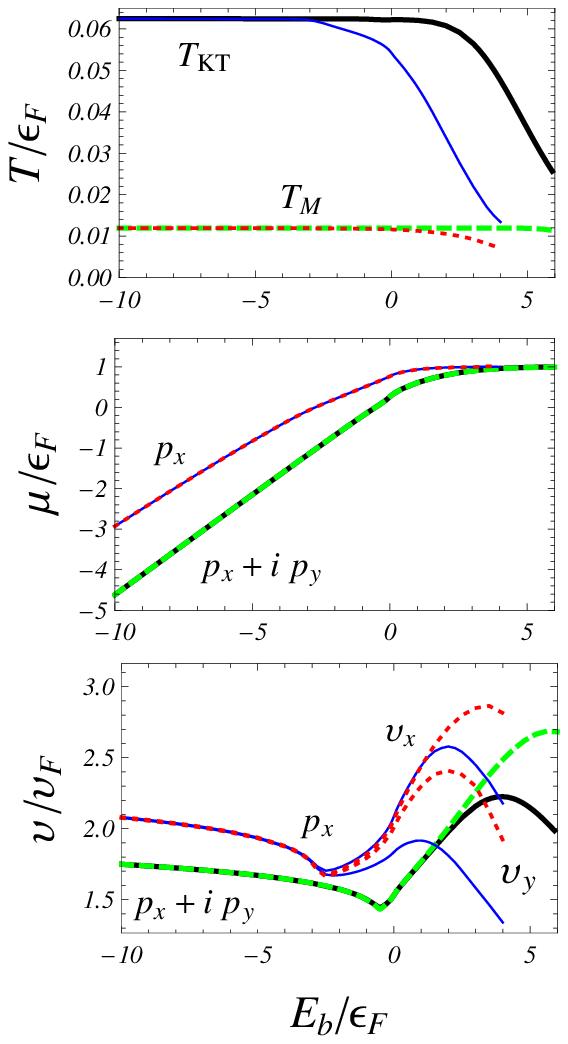}
\put(28,42.5){\includegraphics[scale=.4]{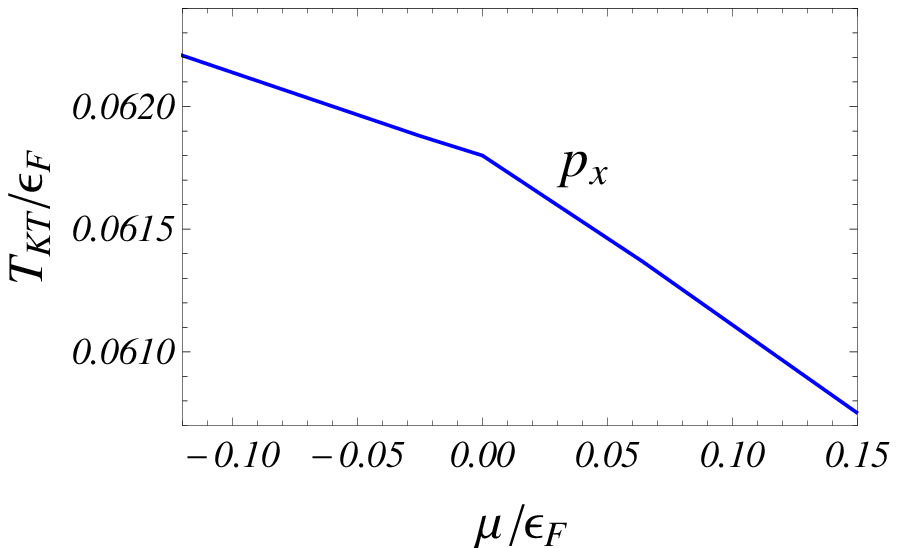}}
\put(13.5,21){\includegraphics[scale=.36]{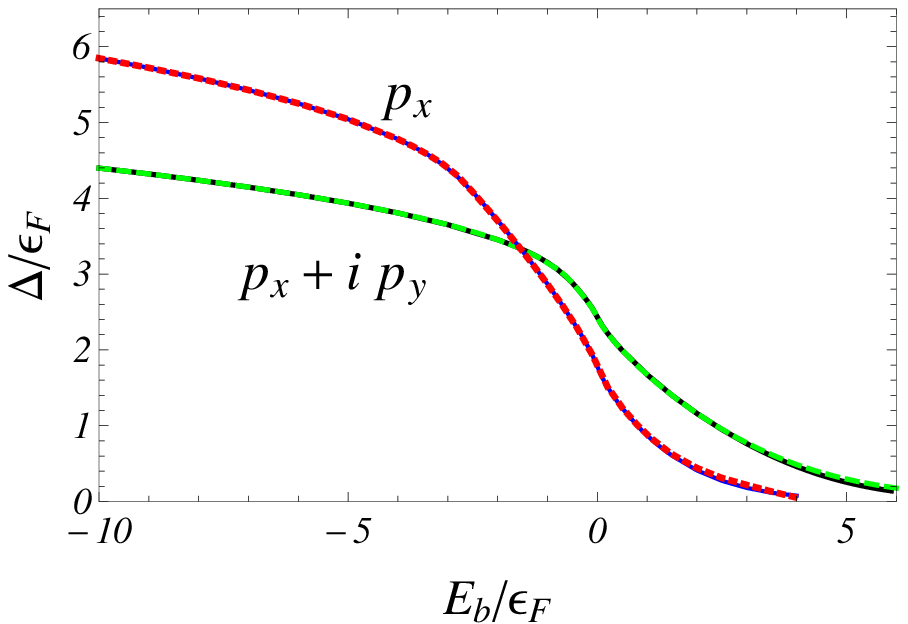}}
\end{overpic}
\caption{(Color online) Upper panel: the transition temperatures, $T_{\rm KT}$ (solid lines) and $T_{\rm M}$ (dashed lines), as functions of the two-body binding energy $E_b$ for the isotropic $p_x+ip_y$ pairing (black and green thick lines) and anisotropic $p_x$ pairing (blue and red thin lines). Lower panels: the chemical potential $\mu$ and sound velocity $\upsilon$ as functions of $E_b$ at $T_{\rm KT}$ and $T_{\rm M}$. The anisotropic velocities for $p_x$ pairing are denoted by $\upsilon_x$ and $\upsilon_y$. The inserts show the zoom-in plots of $T_{\rm KT}$ with respect to $\mu$ around the BCS-BEC transition point $\mu=0$ for the $p_x$ pairing and the order parameter $\Delta$ as a function of $E_b$. The parameters for NSR potential are $k_0=10^{3/2}k_{\rm F}$ and $k_1=10^{1/2}k_{\rm F}$.}\label{plotp}
\end{figure}

To present the numerical results, it is convenient to define the Fermi momentum $k_{\rm F}$ and Fermi energy $\epsilon_{\rm F}$ of a noninteracting Fermi gas, through $n=k_{\rm F}^2/(4\pi)$ and 
$\epsilon_{\rm F}=k_{\rm F}^2/(2m)$. 
The numerical results are shown in Fig.\ref{plotp} in which we plot the transition temperatures $T_{\rm KT}$ and $T_{\rm M}$, the chemical potential, the order parameter, and the sound velocity at $T_{\rm KT}$ and $T_{\rm M}$ as functions of $E_b$. The $E_b$ dependence of $T_{\rm KT}$ clearly shows the BCS-BEC evolution when $E_b$ is tuned from positive to negative values (Note that for $p$-wave pairing in 2D, an attractive potential does not necessarily lead to a bound state; when $E_b>0$ the two-fermion state is a scattering state.). The chemical potential at $T_{\rm KT}$ and $T_{\rm M}$ are almost the same for a given $p$-wave pairings; similarly, the order parameter at $T_{\rm KT}$ and $T_{\rm M}$ are also almost the same. In the deep BEC region where $E_b<0$ with a large magnitude, the transition temperatures $T_{\rm KT}$ and $T_{\rm M}$ are found to be constants $T_{\rm KT}\simeq0.0625\epsilon_{\rm F}$ and $T_{\rm M}\simeq{(0.6/\pi)}T_{\rm KT}$ which are comparable to the $s$-wave pairing case~\cite{Botelho2006}. Besides, the anisotropy in the sound velocity disappears for $p_x$ pairing in deep BEC region as illuminated in the plot of the sound velocities $\upsilon_x$ and $\upsilon_y$, because the basic degrees of freedom are compactly bound bosons now and the Yoshida term in Eq.~(\ref{rhop}) is suppressed.

One interesting feature we observe is that there are non-analytic behavior at the BCS-BEC transition point $\mu=0$. We can see this most clearly from the sound velocity. For other values of $\mu$, the KT and VAL melting transitions are always analytic and smooth. To illuminate this more explicitly, we show the results for $T_{\rm KT}$ around the region $\mu\sim0$ in the inserted figure for the anisotropic $p_x$ pairing, which is more obvious than the isotropic $p_x+ip_y$ pairing. In order to understand the non-analyticity, we explore the properties of the most relevant quantity $\zeta_0^{\rm p}$ around $\mu=0$. The first two derivatives of $\zeta_0^{\rm p}$ with respect to $\mu$ are given by
\begin{eqnarray}
{\partial\zeta_0^{\rm p}\over\partial\mu}&=&\int\!\!\! {d^2{\bf k}\over(2\pi)^2}{\xi_{\bf k}|\Gamma^{\rm p}({\bf k})|^2\over8E_{\bf k}^4}\Bigg[{3\tanh\left({E_{\bf k}\over2T}\right)\over E_{\bf k}}-{{\rm sech}^2\left({E_{\bf k}\over2T}\right)\over2T}\Bigg],\\
{\partial^2\zeta_0^{\rm p}\over\partial\mu^2}&=&\int\!\!\! {d^2{\bf k}\over(2\pi)^2}\frac{|\Gamma^{\rm p}({\bf k})|^2}{8E_{\bf k}^5}\Bigg[{3\Big(5\xi_{\bf k}^2-E_{\bf k}^2\Big)\over E_{\bf k}^2}\tanh\left({E_{\bf k}\over2T}\right)-{|\Delta_{\bf k}^{\rm p}|^2\over2TE_{\bf k}}\nonumber\\
&&\!\!\!\!\times{\rm sech}^2\left({E_{\bf k}\over2T}\right)-{\xi_{\bf k}^2\over2T^2}\tanh\left({E_{\bf k}\over2T}\right){\rm sech}^2\left({E_{\bf k}\over2T}\right)\Bigg].\label{zetap2}
\end{eqnarray}
For small $\mu\rightarrow0^+$, $\partial\zeta_0^{\rm p}/\partial\mu$ is finite but 
\begin{eqnarray}
{\partial^2\zeta_0^{\rm p}\over\partial\mu^2}\sim {{1}\over{T\Delta^4}}\ln{{\mu}\over{\Delta}}.
\end{eqnarray}
As $\zeta_0^{\rm p}$ appears in $n$ and $\rho_{\rm i}^{\rm p}$, this shows that the higher order derivatives of $n$ and $\rho_{\rm i}^{\rm p}$ with respect to $\mu$ is not analytic at the point where $\mu=0$. Also, $T_{\rm KT}$, $T_{\rm M}$, and the sound velocity $\upsilon$ are all non-analytic at the point where $\mu=0$. But we note that the temperature effect weakens the non-analyticities as can be seen from the above equations. Thus, the BCS-BEC evolution in $p$-wave pairing system is actually a phase transition although there is no change of symmetry across the transition. 

\begin{figure}[!htb]
\centering
\includegraphics[width=0.45\textwidth]{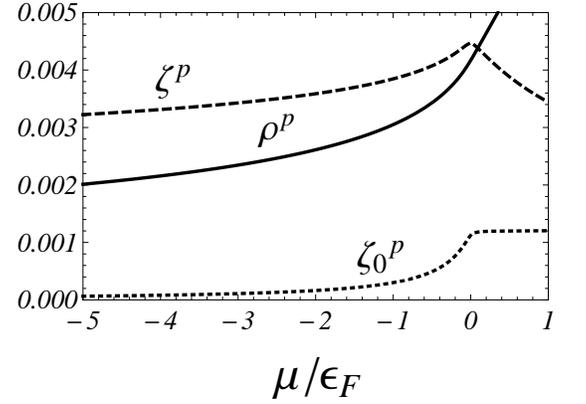}
\caption{The behavior of the quantities $\zeta_0^{\rm p},\zeta^{\rm p}$, and $\rho^{\rm p}$ with respect to the chemical potential $\mu$ for the $p_x+ip_y$ pairing. These quantities have been scaled by proper constants so that they are dimensionless in the plot. In the calculations we choose $T=0.06\epsilon_{\rm F}$ and $\Delta=4\epsilon_{\rm F}$. The parameters for the NSR potential are the same as used in Fig.~\ref{plotp}.}\label{Pvs}
\end{figure}

The sound velocities behave non-monotonically versus $E_b$ and we will analyze it in more detail. For the anisotropic $p_x$ pairing, the sound velocities along $x$ and $y$ directions split in the BCS region 
($E_b$ positive and large) and merge into a single curve in the deep BEC region. For the isotropic $p_x+ip_y$ pairing, we plot the relevant functions $\zeta_0^{\rm p},\zeta^{\rm p}$ and $\rho^{\rm p}$ versus the chemical potential $\mu$ in Fig.\ref{Pvs} to understand the extremas in the sound velocities. As can be seen, the term $B_{11}^2/A_{12}$ dominates $\zeta_0^{\rm p}$ at low temperature which indicates the importance of the $\sigma$ component in $\theta$ mode and the increasing feature of the sound velocities in the BCS region is due to the fast decreasing of $\zeta^{\rm p}$. The sound velocity decreases in the BCS regime with large $E_b$ where $\Delta$ is small. This interesting non-monotonic behavior of the sound velocity may be used to probe the BCS-BEC transition in Fermi gases with $p$-wave pairing.

\section{$d$-wave pairing in spin-$1/2$ fermi gases}\label{chapterd}
\subsection{Formalism in Gaussian approximation}\label{sectiondA}
We now consider a spin-$1/2$ Fermi gas or a two-component Fermi gas with a $d$-wave interaction between the unlike spin components. In this case, Fermi surface mismatch between different spin components can be introduced through Zeeman effect
induced by a magnetic field~\cite{Sarma,LO,FF,Takada}, through imbalance spin populations~\cite{Zwierlein2006,Partridge2006,Sheehy2006}, or through spin-orbit coupling~\cite{Zhai:2014gna}. The Hamiltonian density can be written as~\cite{Botelho2005b}
\begin{eqnarray}
{\cal H}&=&\sum_{{\bf k},s=\uparrow\downarrow}\xi_{{\bf k}s}^{\phantom{\dag}}\psi_{{\bf k},s}^\dagger\psi_{{\bf k},s}^{\phantom{\dag}}
+\sum_{{\bf k,k',q}}V_{{\bf kk'}}^{\rm d}b_{{\bf kq}}^\dagger b_{{\bf k'q}}^{\phantom{\dag}},
\end{eqnarray}
where $\psi_{{\bf k},s}$ represents the fermion annihilation operator with spin $s=\uparrow,\downarrow$, $b_{{\bf kq}}=\psi_{{\bf -k+q}/2,\downarrow}\psi_{{\bf k+q}/2,\uparrow}$ and $\xi_{{\bf k}s}=\xi_{{\bf k}}-s\delta\mu$. Here and in the following, $s=+$($-$) for the spin $\uparrow$($\downarrow$) when we use . For the sake of simplicity, we consider a separable $d$-wave interaction potential~\cite{Cao2013}:
\begin{eqnarray}
V_{{\bf kk'}}^{\rm d}=-\lambda\Gamma^{\rm d}({\bf k})\Gamma^{\rm d*}({\bf k'}),
\end{eqnarray}
where the gamma functions are defined according to NSR-type potentials~\cite{Botelho2005a}:
\begin{eqnarray}
\Gamma^{\rm d}_s({\bf k})={(k_x+ik_y)^2/k_1^2\over(1+k/k_0)^{5/2}},\ \Gamma^{\rm d}_a({\bf k})={(k_x^2-k_y^2)/k_1^2\over(1+k/k_0)^{5/2}}
\end{eqnarray}
with $s$ and $a$ representing the symmetric (or isotropic) $d_{x^2-y^2}+2id_{xy}$ and asymmetric (or anisotropic) $d_{x^2-y^2}$ pairings, respectively. The form of the denominator is chosen to mimic the amplitude damping for $d$-wave partial potential at large momentum~\cite{Botelho2005a}.

Then, the partition function at finite temperature is given by
\begin{eqnarray}
{\cal Z}\!=\!\!\int\!\!\prod_{s=\uparrow\downarrow} [d{\psi_s^\dagger}][d{\psi_s}^{\phantom{\dag}}]\exp\left\{-\!\!\int_0^\beta \!\!\!d\tau\Bigg(\sum_{{\bf k},s=\uparrow\downarrow}
\!\!\psi_{{\bf k},s}^\dagger\partial_\tau\psi_{{\bf k},s}^{\phantom{\dag}}\!+\!{\cal H}\Bigg)\right\}.
\end{eqnarray}
Introducing the auxiliary field $\phi_{\bf q}(\tau)=\lambda\sum_{{\bf k}}\Gamma^{\rm d}({\bf k})b_{{\bf kq}}$ through Hubbard-Stratonovich transformation, the partition function can be rewritten as:
\begin{eqnarray}
{\cal Z}&=&\int  [d{\phi^*}][d{\phi}][d{\Psi^\dagger}][d{\Psi}]\exp\left\{-\int_0^\beta d\tau\Bigg[\sum_{\bf k}{|\phi_{\bf k}(\tau)|^2\over\lambda}\right.\nonumber\\
&&\left.+\sum_{{\bf k,k'}}\Big(\xi_{\bf k}\delta_{\bf k,k'}+\Psi_{{\bf k}}^\dagger G^{-1}_{\bf k,k'}\Psi_{{\bf k'}}^{\phantom{\dag}}\Big)\Bigg]\right\},
\end{eqnarray}
where the fermion field in Nambu-Gor'kov space is $\Psi_{{\bf k}}^\dagger=(\psi_{{\bf k},\uparrow}^\dagger,\psi_{-{\bf k},\downarrow})$. The inverse propagator is then a $2\times2$ matrix which is given by
\begin{eqnarray}
G^{-1}_{\bf k,k'}(\tau)
\!=\!\left(\begin{array}{cc}
\!\!(\partial_\tau\!+\!\xi_{{\bf k}\uparrow})\delta_{\bf k,k'}&\!\!-\phi_{\bf k-k'}(\tau)\Gamma^{\rm d}({\bf k+k'\over2})\\
\!\!-\phi_{\bf -k+k'}^*(\tau)\Gamma^{{\rm d}*}({\bf k+k'\over2})&\!\!(\partial_\tau\!-\!\xi_{{\bf k}\downarrow})\delta_{\bf k,k'}
\end{array}\right).
\end{eqnarray}
Integrating out the fermion degrees of freedom, we can get a bosonic version of partition function
\begin{eqnarray}
{\cal Z}=\int [d{\phi^*}][d{\phi}]~e^{-{\cal S}_{\rm eff}^{\rm d}[\phi^*,\phi]},
\end{eqnarray}
with the effective action
\begin{eqnarray}
{\cal S}_{\rm eff}^{\rm d}=\!\int_0^\beta\! d\tau\Bigg[\sum_{\bf k}\!{|\phi_{\bf k}(\tau)|^2\over\lambda}
+\sum_{{\bf k,k'}}\Big(\xi_{\bf k}\delta_{\bf k,k'}-\text{Tr}\ln G^{-1}_{\bf k,k'}\Big)\Bigg],
\end{eqnarray}
where the trace is taken over imaginary time, momentum and Nambu-Gorkov spaces.  

To proceed, we decompose the auxiliary field $\phi_{\bf q}(\tau)$ into its mean-field and fluctuation parts,
\begin{eqnarray}
\phi_{\bf q}(\tau)=\Delta\delta_{\bf q,0}+\hat\phi_{\bf q}(\tau).
\end{eqnarray}
The effective action can be evaluated in powers of the fluctuation $\hat\phi_{\bf q}(\tau)$, i.e., ${\cal S}_{\rm eff}^{\rm d}={\cal S}_0^{\rm d}+{\cal S}_2^{\rm d}+\cdots$. The leading-order term ${\cal S}_0^{\rm d}$ represents the mean-field contribution. The Gaussian term ${\cal S}_2^{\rm p}$  represents the collective modes.

\subsubsection{Mean field approximation}\label{sectiond1}
The mean-field effective potential can be obtained in a way parallel to the $p$-wave pairing case. We obtain
\begin{eqnarray}
S_0^{\rm d}(\Delta)\!\!\!&=&\!\!\!\beta S\Bigg[{\Delta^2\over\lambda}\!+\int\!\!\! {d^2{\bf k}\over(2\pi)^2}\xi_{\bf k}\!-\!{T\over2}\!\sum_{n}\!\int\!\!\! {d^2{\bf k}\over(2\pi)^2}\ln \text{Det}{\cal G}^{-1}_{\bf k}(i\omega_n)\Bigg]\nonumber\\
&=&\!\!\!{\beta S}\Bigg\{\!{\Delta^2\over\lambda}\!-\!\!\int\!\!\!{d^2{\bf k}\over(2\pi)^2}\Big[E_{\bf k}\!-\!\xi_{\bf k}\!+\!\!\sum_{s=\pm}\!T\ln\big(1\!+\!e^{-{E_{\bf k}+s\delta\mu\over T}}\big)\Big]\Bigg\},
\end{eqnarray}
where the dispersion is $E_{\bf k}=(\xi_{{\bf k}}^{2}+|\Delta_{\bf k}^{\rm d}|^2)^{1/2}$ with the gap function $\Delta_{\bf k}^{\rm d}=\Delta\Gamma^{\rm d}({\bf k})$. The inverse fermion propagator reads
\begin{eqnarray}
{\cal G}^{-1}_{\bf k}(i\omega_m)=\left(\begin{array}{cc}
(i\omega_n+\xi_{{\bf k}\uparrow})&-\Delta_{\bf k}^{\rm d}\\
-\Delta_{\bf k}^{{\rm d}*}&(i\omega_n-\xi_{{\bf k}\downarrow})
\end{array}\right).
\end{eqnarray}
The saddle point condition $\partial S_0^{\rm d}(\Delta)/\partial\Delta=0$ gives the gap equation for the order parameter $\Delta$, 
\begin{eqnarray}\label{gapd}
{2\over\lambda}=\sum_{s=\pm}\int {d^2{\bf k}\over(2\pi)^2}{|\Gamma^{\rm d}({\bf k})|^2\over 2E_{\bf k}}\tanh\left({E_{\bf k}+s\delta\mu\over2T}\right).
\end{eqnarray}
The number density can be obtained through the thermodynamic relation $n=-(\partial S_{\rm eff}^{\rm d}(\Delta)/\partial\mu)/\beta S$. We obtain
\begin{eqnarray}\label{numberd0}
n_0&\equiv&\sum_{s=\pm}\int{d^2{\bf k}\over(2\pi)^2}n_0({\bf k},s)\nonumber\\
&=&{1\over2}\sum_{s=\pm}\int{d^2{\bf k}\over(2\pi)^2}\Bigg[1-{\xi_{{\bf k}}\over E_{\bf k}}\tanh\left({E_{\bf k}+s\delta\mu\over2T}\right)\Bigg].
\end{eqnarray}

Similar to the $p$-wave pairings, the interaction strength $\lambda$ can be physically characterized
by the two-body binding energy $E_b$ in vacuum~\cite{Botelho2005a}:
\begin{eqnarray}
{1\over\lambda}=\int{d^2{\bf k}\over(2\pi)^2}{|\Gamma^{\rm d}({\bf k})|^2\over2\epsilon_{{\bf k}}-E_b}.
\end{eqnarray}
The weak and strong attraction limits correspond to
$E_b\rightarrow+\infty$ and $E_b\rightarrow-\infty$, respectively.

\subsubsection{Gaussian fluctuation and Goldstone mode}\label{sectiond2}
Similar to the $p$-wave pairing case, the the effective action for the collective modes can be evaluated as
\begin{eqnarray}
{\cal S}_2^{\rm d}&=&\sum_{{\bf q},n}\Bigg\{{|\hat\phi_{\bf q}(i\nu_n)|^2\over\lambda}+{T\over2S}\!\sum_{{\bf k},m}{\rm tr}~\Big[{\cal G}_{\bf k-q/2}(i\omega_m) \nonumber\\
&&\times~\Phi_{\bf -q}(-i\nu_n){\cal G}_{\bf k+q/2}(i\omega_m+i\nu_n)\Phi_{\bf q}(i\nu_n)\Big]\Bigg\},
\end{eqnarray}
where the fermion propagator and the matrix $\Phi$ are given by
\begin{eqnarray}
{\cal G}_{\bf k}(i\omega_n)&=&{1\over(i\omega_n-\delta\mu)^2-E_{{\bf k}}^2}\left(\begin{array}{cc}
(i\omega_n-\xi_{{\bf k}\downarrow})&\Delta_{\bf k}^{\rm d}\\
\Delta_{\bf k}^{\rm d*}&(i\omega_n+\xi_{{\bf k}\uparrow})
\end{array}\right),\nonumber\\
\Phi_{\bf q}(i\nu_n)&=&\left(\begin{array}{cc}
0&\!\!-\hat\phi_{\bf q}(i\nu_n)\Gamma^{\rm d}({\bf k})\\
-\hat\phi_{\bf -q}^*(-i\nu_n)\Gamma^{\rm d*}({\bf k})&0
\end{array}\right).
\end{eqnarray}
After some algebra, ${\cal S}_2^{\rm d}$ can be written in a compact form 
\begin{eqnarray}
{\cal S}_2^{\rm d}=\frac{1}{2}\sum_{{\bf q},n}\left(\begin{array}{cc}
\hat{\phi}_{\bf q}^*(i\nu_n) & \hat{\phi}_{-{\bf q}}(-i\nu_n)\end{array}\right) M({\bf q},i\nu_n)\left(\begin{array}{cc} \hat{\phi}_{\bf q}(i\nu_n)\\
\hat{\phi}_{-{\bf q}}^*(-i\nu_n)\end{array}\right),
\end{eqnarray}
where the inverse boson propagator $M({\bf q},i\nu_n)$ takes the form
\begin{eqnarray}
M({\bf q},i\nu_n)=\left(\begin{array}{cc}M_{11}({\bf q},i\nu_n)&M_{12}({\bf q},i\nu_n)\\
M_{21}({\bf q},i\nu_n)& M_{22}({\bf q},i\nu_n)\end{array}\right).
\end{eqnarray}
The matrix elements of $M$ are given by
\begin{eqnarray}
M_{11}&=&{1\over\lambda}+{T\over S}\sum_{{\bf k},m}{\cal G}_{\bf k-q/2}^{11}(i\omega_m){\cal G}_{\bf k+q/2}^{22}(i\omega_m+i\nu_n)|\Gamma^{\rm d}({\bf k})|^2,\nonumber\\
M_{22}&=&{1\over\lambda}+{T\over S}\sum_{{\bf k},m}{\cal G}_{\bf k-q/2}^{22}(i\omega_m){\cal G}_{\bf k+q/2}^{11}(i\omega_m+i\nu_n)|\Gamma^{\rm d}({\bf k})|^2,\nonumber\\
M_{12}&=&{T\over S}\sum_{{\bf k},m}{\cal G}_{\bf k-q/2}^{12}(i\omega_m){\cal G}_{\bf k+q/2}^{12}(i\omega_m+i\nu_n)[\Gamma^{\rm d*}({\bf k})]^2,\nonumber\\
M_{21}&=&{T\over S}\sum_{{\bf k},m}{\cal G}_{\bf k-q/2}^{21}(i\omega_m){\cal G}_{\bf k+q/2}^{21}(i\omega_m+i\nu_n)[\Gamma^{\rm d}({\bf k})]^2.
\end{eqnarray}
It is easy to prove that these matrix elements satisfy 
\begin{eqnarray}
M_{11}^*({\bf q},i\nu_n)=M_{22}({\bf q},i\nu_n),\ \ M_{12}^*({\bf q},i\nu_n)=M_{21}({\bf q},i\nu_n).
\end{eqnarray}
Completing the summation over the fermion Matsubara frequency $i\omega_m$ we obtain
\begin{widetext}
	\begin{eqnarray}\label{matrixd}
\!\!\!	M_{11}\!\!&=&\!\!{1\over\lambda}\!+\!\sum_{s=\pm}\!\int\!\!\! {d^2{\bf k}\over(2\pi)^2}{|\Gamma^{\rm d}({\bf k})|^2\over2}\left[\Bigg({u_-^2\upsilon_+^2\over i\nu_n\!+\!(E_+\!-\!E_-)}\!-\!{u_+^2\upsilon_-^2\over i\nu_n-(E_+-E_-)}\Bigg)(f_+^s\!-\!f_-^s)\!+\!\Bigg({u_+^2u_-^2\over i\nu_n\!-\!(E_+\!+\!E_-)}\!-\!{\upsilon_+^2\upsilon_-^2\over i\nu_n+(E_+\!+\!E_-)}\Bigg)(1\!-\!f_+^s\!-\!f_-^s)\right],\nonumber\\
\!\!\!	M_{12}\!\!&=&\!\!-\!\sum_{s=\pm}\!\int\!\!\! {d^2{\bf k}\over(2\pi)^2}{[\Gamma^{{\rm d}*}({\bf k})]^2\over8E_+E_-}\left[\Bigg({\Delta_{{\bf k}-{\bf q}/2}^{\rm d}\Delta_{{\bf k}+{\bf q}/2}^{\rm d}\over i\nu_n\!-\!(E_+\!-\!E_-)}\!-\!{\Delta_{{\bf k}-{\bf q}/2}^{\rm d}\Delta_{{\bf k}+{\bf q}/2}^{\rm d}\over i\nu_n+(E_+-E_-)}\Bigg)(f_+^s\!-\!f_-^s)\!+\!\Bigg({\Delta_{{\bf k}-{\bf q}/2}^{\rm d}\Delta_{{\bf k}+{\bf q}/2}^{\rm d}\over i\nu_n\!-\!(E_+\!+\!E_-)}\!-\!{\Delta_{{\bf k}-{\bf q}/2}^{\rm d}\Delta_{{\bf k}+{\bf q}/2}^{\rm d}\over i\nu_n+(E_+\!+\!E_-)}\Bigg)(1\!-\!f_+^s\!-\!f_-^s)\right],
	\end{eqnarray}
\end{widetext}
where the Fermi-Dirac distribution function is given by $f_\pm^s=\Big(1+e^{(E_\pm+s\delta\mu)/T}\Big)^{-1}$. Again, we note that the terms proportional to $f_+^s-f_-^s$ corresponds to the Landau damping effect, which vanish for the balanced case $\delta\mu=0$ when $T\rightarrow0$.

It is more physical to decompose the collective mode $\hat\phi(x)$ into a sum of real and imaginary parts, that is, $\hat\phi(x)=\sigma(x)+i\pi(x)$. Then  Gaussian fluctuation part of the effective action can be reexpressed as
\begin{eqnarray}
{\cal S}_2^{\rm d}={1\over2}\sum_{{\bf q},n}\left(\begin{array}{cc}\sigma_{\bf q}^*(i\nu_n)&\pi_{\bf q}^*(i\nu_n)\end{array}\right)\Pi({\bf q},i\nu_n)\left(\begin{array}{c}\sigma_{\bf q}(i\nu_n)\\ \pi_{\bf q}(i\nu_n)\end{array}\right),
\end{eqnarray}
where the effective inverse boson propagator is
\begin{eqnarray}
\Pi\!=\!\left(\begin{array}{cc}
\!\!(M_{11}\!+\!M_{12}\!+\!M_{21}\!+\!M_{22})\!&\!\!i(\!-\!M_{11}\!-\!M_{12}\!+\!M_{21}\!+\!M_{22})\!\!\!\\
\!\!i(M_{11}\!-\!M_{12}\!+\!M_{21}\!-\!M_{22})\!&\!\!(M_{11}\!-\!M_{12}\!-\!M_{21}\!+\!M_{22})\!\!\!
\end{array}\right).
\end{eqnarray}
Thus, all the matrix elements of $\Pi$ are real and the propagators
of independent collective modes can be obtained through the
diagonalization and we find
\begin{eqnarray}
{\cal D}_{\theta/\eta}^{-1}({\bf q},i\nu_n)=M_{11}+M_{22}\mp\sqrt{(M_{11}-M_{22})^2+4M_{12}M_{21}}.
\end{eqnarray}
It can be verified that ${\cal D}_{\theta}^{-1}(0,0)=0$ which shows $\theta$ to be the Goldstone mode with the following mixing of $\sigma$ and $\pi$ components:
\begin{eqnarray}
\theta_{\bf q}(i\nu_n)={\cal C}\Big[{\cal D}_{\theta}^{-1}({\bf q},i\nu_n)\sigma_{\bf q}(i\nu_n)+{\cal D}_{\eta}^{-1}({\bf q},i\nu_n)\pi_{\bf q}(i\nu_n)\Big],
\end{eqnarray}
where ${\cal C}$ is a normalization coefficient.

The approach for the KT and VAL transitions are the same as we adopted for the $p$-wave pairing. At small energy and momentum, 
the propagator of the gapless $\theta$ mode can be expressed as
\begin{eqnarray}\label{GM-EXP-D}
{\cal D}_\theta^{-1}({\bf q},i\nu_n)=-\zeta^{\rm d}(i\nu_n)^2+{1\over 4m\Delta^2}\rho^{\rm d}{\bf q}^2,
\end{eqnarray}
where we can show that the stiffness $\rho^{\rm d}$ is isotropic for both isotropic and anisotropic $d$-wave pairings.
To compute the coefficients $\zeta^{\rm d}$ and $\rho^{\rm d}$, we make the low-energy expansion of $M_{\rm ij}$ (${\rm i,j}=1,2$) to the quadratic order in frequency and momentum, 
\begin{eqnarray}
M_{\rm ij}({\bf q},i\nu_n)=A_{\rm ij}+i\nu_nB_{\rm ij}+(i\nu_n)^2C_{\rm ij}+D_{\rm ij}^xq_x^2+D_{\rm ij}^yq_y^2.
\end{eqnarray}
The Landau damping problem still exists here.  We again neglect the damping of collective modes and perform this expansion. 
The expansion coefficients read
\begin{eqnarray}\label{ABCd}
&&A_{11}=A_{22}={1\over\lambda}-{1\over S}\sum_{{\bf k},s=\pm}(E_{\bf k}^2+\xi_{\bf k}^2){|\Gamma^{\rm d}({\bf k})|^2\over8E_{\bf k}^3}\tanh\left({E_{\bf k}+s\delta\mu\over2T}\right),\nonumber\\
&&B_{11}=-B_{22}=-{1\over S}\sum_{{\bf k},s=\pm}{\xi_{\bf k}|\Gamma^{\rm d}({\bf k})|^2\over8E_{\bf k}^3}\tanh\left({E_{\bf k}+s\delta\mu\over2T}\right),\nonumber\\
&&C_{11}=C_{22}=-{1\over S}\sum_{{\bf k},s=\pm}(E_{\bf k}^2+\xi_{\bf k}^2){|\Gamma^{\rm d}({\bf k})|^2\over32E_{\bf k}^5}\tanh\left({E_{\bf k}+s\delta\mu\over2T}\right),\nonumber\\
&&A_{12}=A_{21}={1\over S}\sum_{{\bf k},s=\pm}{|\Delta_{\bf k}^{\rm d}|^2|\Gamma^{\rm d}({\bf k})|^2\over8E_{\bf k}^3}\tanh\left({E_{\bf k}+s\delta\mu\over2T}\right),\nonumber\\
&&B_{12}=B_{21}=0,\nonumber\\
&&C_{12}=C_{21}={1\over S}\sum_{{\bf k},s=\pm}{|\Delta_{\bf k}^{\rm d}|^2|\Gamma^{\rm d}({\bf k})|^2\over32E_{\bf k}^5}\tanh\left({E_{\bf k}+s\delta\mu\over2T}\right).
\end{eqnarray}
The coefficients $D_{\rm ij}$ is again rather lengthy and we show the combined quantities 
\begin{eqnarray}
\rho_i^{\rm d}=4m\Delta^2(D_{11}^i+D_{22}^i-D_{12}^i-D_{21}^i), \ (i=x,y)
\end{eqnarray}
which are exactly the superfluid densities along the $x$ and $y$ directions. After a lengthy calculation we obtain
\begin{eqnarray}\label{rhod}
\rho_i^{\rm d}=\sum_{s=\pm}\int\!\!\! {d^2{\bf k}\over(2\pi)^2}\left[n_0({\bf k},s)-\frac{k_i^2}{4mT}{\rm sech}^2\left({E_{\bf k}+s\delta\mu\over2T}\right)\right].
\end{eqnarray}
Compared to $p$-wave pairing, the superfluid density is isotropic for both $d_{x^2-y^2}+2id_{xy}$ and $d_{x^2-y^2}$ pairings at any temperature. We have $\rho_x^{\rm d}=\rho_y^{\rm d}=\rho^{\rm d}$.

Finally, the low-energy behavior of the $\theta$-mode or the Goldstone mode can be given by Eq. (\ref{GM-EXP-D})
where $\zeta^{\rm d}=\zeta_0^{\rm d}+B_{11}^2/A_{12}$ with $\zeta_0^{\rm d}=-2(C_{11}-C_{12})$. The sound velocity is given by 
\begin{eqnarray}
\upsilon^{\rm d}=\sqrt{{\rho^{\rm d}\over 4m\Delta^2\zeta^{\rm d}}}.
\end{eqnarray}
As mentioned in the $p$-wave case, the coupling term $B_{11}^2/A_{12}$ ensures that we recover the correct sound velocity in the BCS-BEC evolution.
The Goldstone mode contribution to the thermodynamic potential can be given by
\begin{eqnarray}
\Omega_2^{\rm d}=\int{d^2{\bf q}\over(2\pi)^2}T\ln\Big(1-e^{-\varepsilon_{\bf q}/T}\Big)=-{\zeta(3)T^3\over2\pi(\upsilon^{\rm d})^2},
\end{eqnarray}
where the dispersion relation is $\varepsilon_{\bf q}=\upsilon^{\rm d}|{\bf q}|$. At finite temperature, we take into account the fluctuation contribution to the number density. 
The total fermion density $n$ is given by
\begin{eqnarray}\label{numberd}
n=n_0-{\partial \Omega_2^{\rm d}\over{\partial\mu}}=n_0-{\zeta(3)T^3\over\pi(\upsilon^{\rm d})^3}{\partial\upsilon^{\rm d}\over\partial\mu},
\end{eqnarray}
which reduces to the mean-field result $n=n_0$ at zero temperature.

\subsection{KT and VAL melting transitions}\label{sectiondB}
In the following, we explore the feature of KT and VAL transitions in this spin-$1/2$ Fermi system with $d$-wave paring. The KT and VAL melting temperatures are both directly related to the stiffness $J=\rho^{\rm d}/(4m)$ in the following way~\cite{BKT,Nelson1979,Young,Botelho2006}:
\begin{eqnarray}\label{TCd}
T_{\rm KT}={\pi\over2}J^{\rm d}(T_{\rm KT}),\ T_{\rm M}=0.3J^{\rm d}(T_{\rm M}).
\end{eqnarray}
The transition temperatures $T_{\rm KT}$ and $T_{\rm M}$ can be determined by solving the gap equation (\ref{gapd}), the number equation (\ref{numberd}), and critical temperature equation (\ref{TCd})  self-consistently.  In the following we will consider balanced ($\delta\mu=0$) and imbalanced ($\delta\mu\neq0$) systems. To present the numerical results, we define the Fermi momentum $k_{\rm F}$ and Fermi energy $\epsilon_{\rm F}$ of a noninteracting balanced Fermi gas, through $n=k_{\rm F}^2/(2\pi)$ and $\epsilon_{\rm F}=k_{\rm F}^2/(2m)$. 

\subsubsection{Balanced Fermi gases}\label{sectiond3}
\begin{figure}[!htb]
\centering
\begin{overpic}
[scale=1.5]{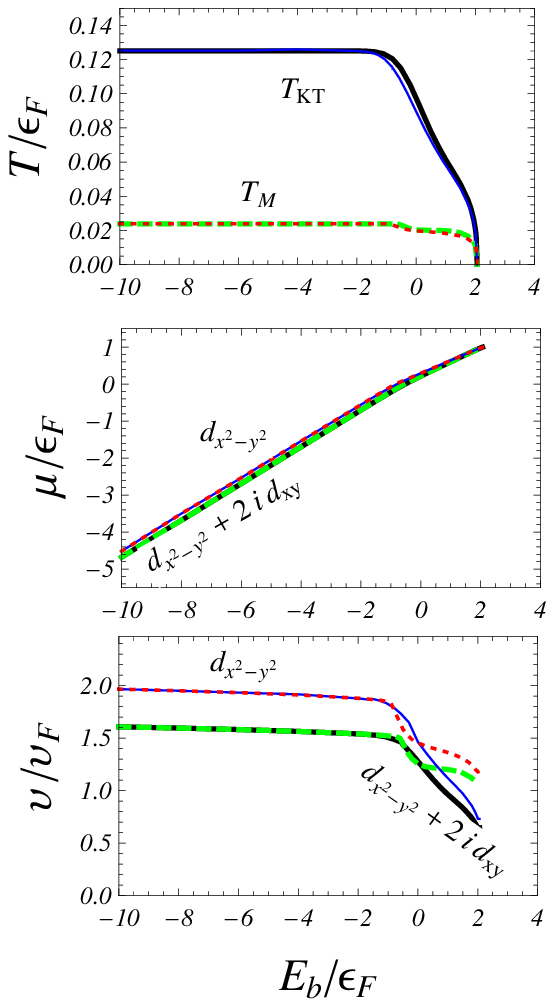}
\put(30,43){\includegraphics[scale=.36]{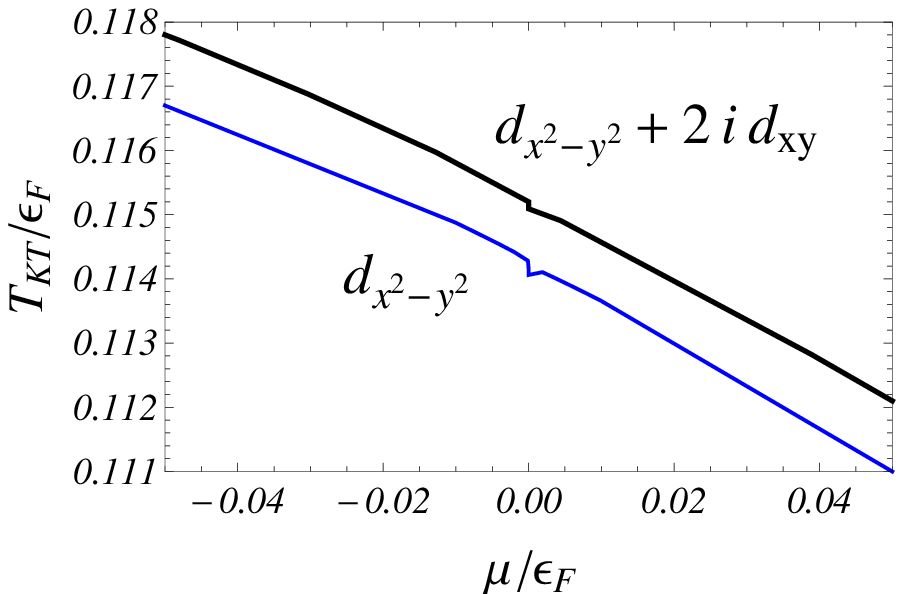}}
\put(13,14){\includegraphics[scale=.36]{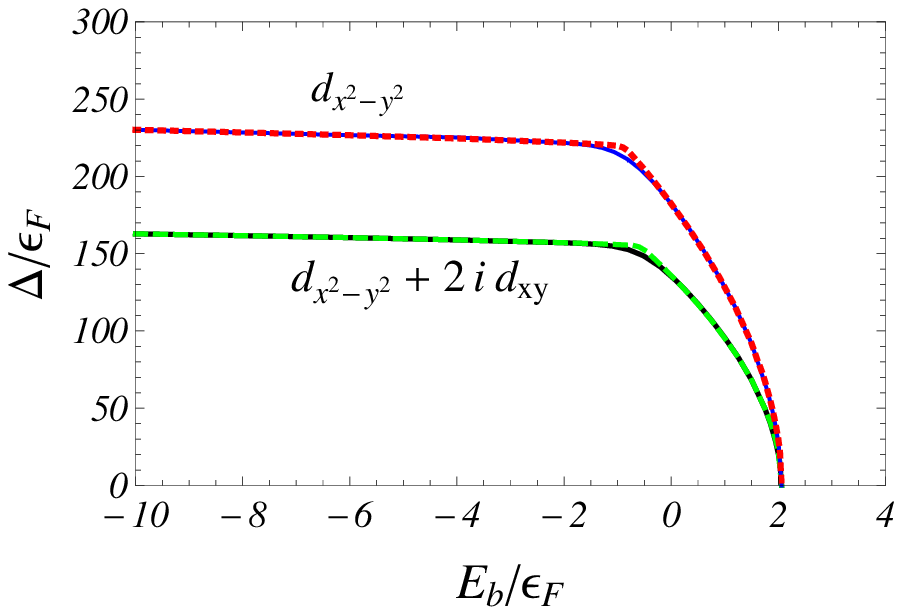}}
\end{overpic}
\caption{(Color online) Upper panel: the transition temperatures, $T_{\rm KT}$ (solid lines) and $T_{\rm M}$ (dashed lines), as functions of the two-body binding energy $E_b$ for the isotropic $d_{x^2-y^2}+2id_{xy}$ pairing (black and green thick lines) and the anisotropic $d_{x^2-y^2}$ pairing (blue and red thin lines). Lower panels: the chemical potential $\mu$ and sound velocity $\upsilon$ as functions of $E_b$ at $T_{\rm KT}$ and $T_{\rm M}$. The inserts show the zoom-in plots of $T_{\rm KT}$ with respect to $\mu$ around the BCS-BEC transition point $\mu=0$ and the the order parameter $\Delta$ as a function of $E_b$. The parameters for NSR potential are $k_0=10^{3/2}k_{\rm F}$ and $k_1=10^{1/2}k_{\rm F}$.}\label{plotd}
\end{figure}

For the balanced system with $\delta\mu=0$, the numerical results are shown in Fig.\ref{plotd}. The transition temperatures $T_{\rm KT}$ and $T_{\rm M}$ approach constants in the deep BEC region, $T_{\rm KT}=0.125\epsilon_{\rm F}$ and $T_{\rm M}={(0.6/\pi)}T_{\rm KT}$, as we found in the $p$-wave pairing case. The chemical potential, the order parameter, and the sound velocity are not sensitive to the temperature. At intermediate and at strong coupling, their values at $T_{\rm M}$ and $T_{\rm KT}$ are almost the same.
.

For the $d$-wave paring case, the non-analyticity is also found at the BCS-BEC transition point $\mu=0$. To see this, we can take the same argument as we gave for the $p$-wave pairing case. We explore the properties of the most relevant quantity $\zeta_0^{\rm d}$ around $\mu=0$. The derivative of $\zeta_0^{\rm d}$ with respect to $\mu$ is given by:
\begin{eqnarray}
{\partial\zeta_0^{\rm d}\over\partial\mu}&=&\!\!\int\!\!\! {d^2{\bf k}\over(2\pi)^2}{\xi_{\bf k}|\Gamma^{\rm d}({\bf k})|^2\over4E_{\bf k}^4}\Bigg[{3\tanh\Big({E_{\bf k}\over2T}\Big)\over E_{\bf k}}-{{\rm sech}^2\Big({E_{\bf k}\over2T}\Big)\over2T}\Bigg],
\end{eqnarray}
which is divergent logarithmically at $\mu=0$. This further induces non-analyticities in $n$, $T_{\rm KT}$, $T_{\rm M}$, etc. In the numerical results shown in Fig.~\ref{plotd}, the non-analyticities are not obvious for the present choice of the parameters $k_0$ and $k_1$; however, we can easily identify the non-analyticity in the inserted figure for $T_{\rm KT}$. 

\begin{figure}[!htb]
\centering
\includegraphics[width=0.45\textwidth]{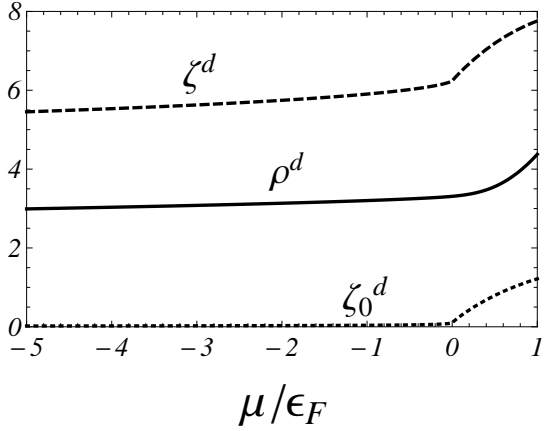}
\caption{The behavior of the quantities $\zeta_0^{\rm d},\zeta^{\rm d}$, and $\rho^{\rm d}$ with respect to the chemical potential $\mu$ for the $d_{x^2-y^2}+2id_{xy}$ pairing. These quantities have been scaled by proper constants so that they are dimensionless in the plot and the magnitudes are all increased by multiplying $10^6$. In the calculations we choose  $T=0.12\epsilon_{\rm F}$ and $\Delta=150\epsilon_{\rm F}$. The parameters for the NSR potential are the same as used in Fig.~\ref{plotd}.}\label{Dvs}
\end{figure}

Unlike the $p$-wave pairing case, the sound velocity for $d$-wave pairing does not show non-monotonicity: it is always a decreasing function when $E_b$ goes from negative to positive. This can be understood from Fig.\ref{Dvs}: As $\zeta^{\rm d}$ always increases faster than $\rho^{\rm d}$ with $\mu$, the sound velocity $\upsilon^{\rm d}$ decreases monotonously with the binding energy $E_b$.

\subsubsection{Mismatched Fermi gases}\label{sectiond4}
In order to study the effect of mismatched Fermi surfaces ($\delta\mu\neq0$) on KT and VAL melting transitions, we choose a fixed binding energy $E_b=-2\epsilon_{\rm F}$ as an example which lies in the BEC region.

\begin{figure}[!htb]
\centering
\begin{overpic}
[scale=1.4]{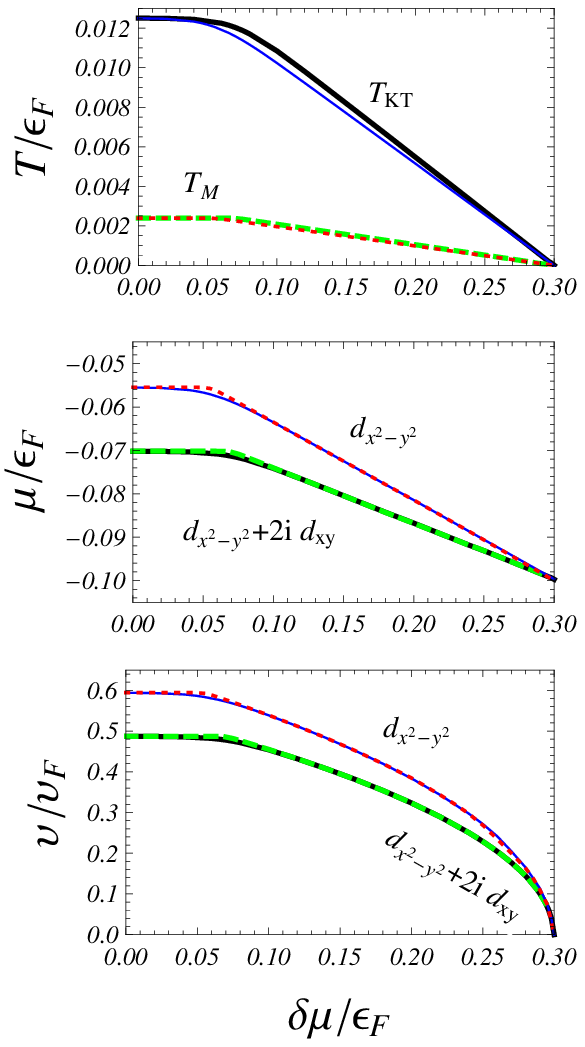}
\put(12.5,11){\includegraphics[scale=.35]{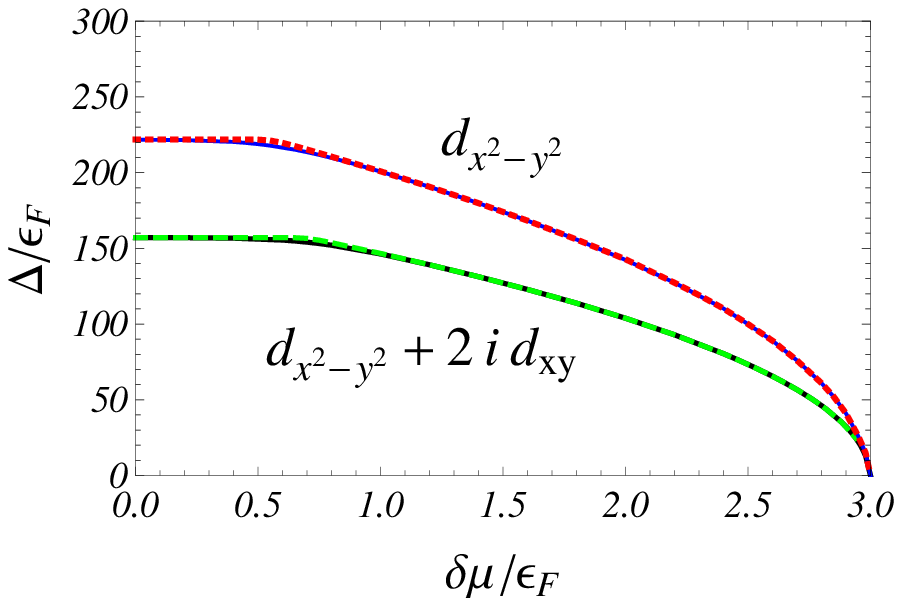}}
\end{overpic}
\caption{(Color online)  Upper panel: the transition temperatures, $T_{\rm KT}$ (solid lines) and $T_{\rm M}$ (dashed lines), as functions of the Zeeman field $\delta\mu$ with a fixed binding energy $E_b=-2\epsilon_{\rm F}$ for the isotropic $d_{x^2-y^2}+2id_{xy}$ pairing (black and green thick lines) and the anisotropic $d_{x^2-y^2}$ pairing (blue and red thin lines). Lower panels: the chemical potential $\mu$ and sound velocity $\upsilon$ as functions of $\delta\mu$ at $T_{\rm KT}$ and $T_{\rm M}$. The insert shows the order parameter $\Delta$ as a function of $\delta\mu$. The parameters for the NSR potential are the same as used in Fig.~\ref{plotd}.}\label{Ddmu}
\end{figure}

We plot the transition temperatures $T_{\rm KT}$ and $T_{\rm M}$, the chemical potentials at $T_{\rm KT}$ and $T_{\rm M}$, the order parameters $\Delta$ at $T_{\rm KT}$ and $T_{\rm M}$, and the sound velocity $\upsilon$  at $T_{\rm KT}$ and $T_{\rm M}$ as functions of $\delta\mu$ in Fig.\ref{Ddmu}. As we expect, all these quantities decreases with $\delta\mu$. For small $\delta\mu$, the decreasing effect is not significant. However, for large $\delta\mu$, they almost linearly decrease with $\delta\mu$ and finally reach a critical point $\delta\mu_c\!\sim\!3\epsilon_{\rm F}$ beyond which the superfluidity is destroyed and the KT and VAL melting transition temperatures both approach zero at this point.

\section{summary}\label{summary}
In this work, the features of the Kosterlitz-Thouless and vortex-antivortex lattice melting transitions are explored in detail for fermionic systems with higher partial wave pairings, including $p$-wave and $d$-wave pairings.  The KT and VAL melting transitions are obtained by studying the low-energy dynamics of the gapless Goldstone mode.  Our approach takes into account both the amplitude and phase modes and can recover the correct sound velocity in the BCS-BEC evolution, which enables us to include correctly the collective modes contribution to the thermodynamics.

The main results in this work can be summarized as follows: \\
\textbf{(a)} For the $p$-wave pairing, we find that the transition temperatures $T_{\rm KT}$ and $T_{\rm M}$ approach constants in the BEC region: $T_{\rm KT}=0.0625\epsilon_{\rm F}$ and $T_{\rm M}={(0.6/\pi)}T_{\rm KT}$. The KT transition temperature is thus reachable in current cold atom experiments. The transition temperatures and the sound velocities are continuous but non-analytic across the BCS-BEC transition point $\mu=0$. For the anisotropic $p_x$ pairing, the sound velocity is anisotropic in BCS region but becomes nearly isotropic in the BEC region. The sound velocity exhibits non-monotonic behavior and may be used to probe the BCS-BEC transition in Fermi gases with $p$-wave pairing. \\
\textbf{(b)} For $d$-wave pairing, the transition temperatures $T_{\rm KT}$ and $T_{\rm M}$ also approach constants in the BEC region: $T_{\rm KT}=0.125\epsilon_{\rm F}$ and $T_{\rm M}={(0.6/\pi)}T_{\rm KT}$. The transition temperatures and sound velocities are noncontinuous across the BCS-BEC transition point $\mu=0$ because of the higher divergence degree~\cite{Cao2013}. Because of the exchange symmetry between $k_x$ and $k_y$, the sound velocity is isotropic even for the anisotropic $d_{x^2-y^2}$ pairing. We find that the effect of mismatched Fermi surfaces also destroys the $d$-wave superfluidity and the associated KT transition.

\acknowledgments
We thank Hui Hu for discussion. This work is supported by the Thousand Young Talents Program of China. G.C. and X.G.H. are also supported by NSFC with Grant No. 11535012 and No. 11675041, and Scientific Research Foundation of State Education Ministry for Returned Scholars. G.C. is also supported by China Postdoctoral Science Foundation with Grant No. KLH1512072. L. H acknowledges the support by NSFC with Grant No. 11775123.

\appendix
\begin{widetext}
\section{The expansion coefficients for the $\theta$ mode}\label{rho}
In order to obtain the analytic form of the stiffness, we expand $M_{11}+M_{22}-M_{12}-M_{21}$ for small ${\bf q}$ at $i\nu_n=0$. We take the $d$-wave pairing case as an example and the $p$-wave pairing case is similar. The relevant term is
\begin{eqnarray}
F({\bf q})&=&\sum_{{\bf k},m}{1\over\Big[(i\tilde\omega_m)^2-E_{\bf k-q/2}^2\Big]\Big[(i\tilde\omega_m)^2-E_{\bf k+q/2}^2\Big]}\Bigg\{2(i\tilde\omega_m-\xi_{\bf k-q/2})(i\tilde\omega_m+\xi_{\bf k+q/2})
|\Gamma^{\rm d}({\bf k})|^2\nonumber\\
&&-\Delta_{\bf k-q/2}\Delta_{\bf k+q/2}[\Gamma^{\rm d*}({\bf k})]^2-\Delta_{\bf k-q/2}^*\Delta_{\bf k+q/2}^*[\Gamma^{\rm d}({\bf k})]^2\Bigg\},
\end{eqnarray}
where $i\tilde\omega_m=i\omega_m-\delta\mu$. The first derivative of $F({\bf q})$ with respect to $q_i\ (i=x,y)$ is
\begin{eqnarray}
\partial_{q_i}F({\bf q})&=&\sum_{{\bf k},m}{1\over\Big[(i\tilde\omega_m)^2-E_{\bf k-q/2}^2\Big]\Big[(i\tilde\omega_m)^2-E_{\bf k+q/2}^2\Big]}\Bigg\{\left[\partial_{k_i}\xi_{\bf k-q/2}(i\tilde\omega_m+\xi_{\bf k+q/2})+\partial_{k_i}\xi_{\bf k+q/2}(i\tilde\omega_m-\xi_{\bf k-q/2})\right]|\Gamma^{\rm d}({\bf k})|^2\nonumber\\
&&+{1\over2}\left(\Delta_{\bf k-q/2}'\Delta_{\bf k+q/2}-\Delta_{\bf k-q/2}\Delta_{\bf k+q/2}'\right)[\Gamma^{{\rm d}*}({\bf k})]^2+{1\over2}\left(\Delta_{\bf k-q/2}^{*'}\Delta_{\bf k+q/2}^*
-\Delta_{\bf k-q/2}^{*}\Delta_{\bf k+q/2}^{*'}\right)[\Gamma^{\rm d}({\bf k})]^2\Bigg\}\nonumber\\
&&-\sum_{{\bf k},m}{1\over2\Big[(i\tilde\omega_m)^2-E_{\bf k-q/2}^2\Big]^2\Big[(i\tilde\omega_m)^2-E_{\bf k+q/2}^2\Big]^2}\Bigg\{\left(E_{\bf k-q/2}^2\right)'\Big[(i\tilde\omega_m)^2-E_{\bf k+q/2}^2\Big]
-\left(E_{\bf k+q/2}^2\right)'\Big[(i\tilde\omega_m)^2-E_{\bf k-q/2}^2\Big]\Bigg\}\nonumber\\
&&\Bigg\{2(i\tilde\omega_m-\xi_{\bf k-q/2})(i\tilde\omega_m+\xi_{\bf k+q/2})|\Gamma^{\rm d}({\bf k})|^2-\Delta_{\bf k-q/2}\Delta_{\bf k+q/2}[\Gamma^{{\rm d}*}({\bf k})]^2
-\Delta_{\bf k-q/2}^*\Delta_{\bf k+q/2}^*[\Gamma^{\rm d}({\bf k})]^2\Bigg\}.
\end{eqnarray}
Here we use the notation $A^\prime=\partial_{q_i}A$. Keeping in mind that $\Big[(i\tilde\omega_m)^2-E_{\bf k-q/2}^2\Big]\Big[(i\tilde\omega_m)^2-E_{\bf k+q/2}^2\Big]$ is an even function of ${\bf q}$, we can evaluate the second derivative of $F({\bf q})$ around ${\bf q}=0$ as
\begin{eqnarray}\label{F2}
\partial_{q_i}^2F({\bf q})|_{\bf q=0}
&=&\sum_{{\bf k},m}{1\over\Big[(i\tilde\omega_m)^2-E_{\bf k}^2\Big]^2}\Bigg\{\Big[\xi_{\bf k}\partial_{k_i}^2\xi_{\bf k}+3(\partial_{k_i}\xi_{\bf k})^2\Big]|\Gamma^{\rm d}({\bf k})|^2+{\Delta^2\over4}\left(|\Gamma^{\rm d}({\bf k})|^4\right)''\Bigg\}\nonumber\\
&&+\sum_{{\bf k},m}{4|\Gamma^{\rm d}({\bf k})|^2\over\Big[(i\tilde\omega_m)^2-E_{\bf k}^2\Big]^3}\Big[\xi_{\bf k}\partial_{k_i}\xi_{\bf k}+{1\over2}\left(|\Delta_{\bf k}|^2\right)'\Big]^2.
\end{eqnarray}
Finally, we complete the Matsubara frequency summation and obtain
\begin{eqnarray}\label{F21}
\partial_{q_i}^2F({\bf q})|_{\bf q=0}&=&\sum_{{\bf k},s=\pm}\Bigg\{\Big[\xi_{\bf k}\partial_{k_i}^2\xi_{\bf k}+3(\partial_{k_i}\xi_{\bf k})^2\Big]|\Gamma^{\rm d}({\bf k})|^2+{\Delta^2\over4}
\left(|\Gamma^{\rm d}({\bf k})|^4\right)''\Bigg\}\left[{\tanh\left({E_{\bf k}^s\over2T}\right)\over8E_{\bf k}^3}-{{\rm sech}^2\left({E_{\bf k}^s\over2T}\right)\over16TE_{\bf k}^2}\right]\nonumber\\
&&+\sum_{{\bf k},s=\pm}\Big[\xi_{\bf k}\partial_{k_i}\xi_{\bf k}+{1\over2}\left(|\Delta_{\bf k}|^2\right)'\Big]^2\frac{|\Gamma^{\rm d}({\bf k})|^2}{8E_{\bf k}^3}
\left[-{3\tanh\left({E_{\bf k}^s\over2T}\right)\over E_{\bf k}^2}+{3{\rm sech}^2\left({E_{\bf k}^s\over2T}\right)\over 2TE_{\bf k}}+{\tanh\left({E_{\bf k}^s\over2T}\right){\rm sech}^2\left({E_{\bf k}^s\over2T}\right)\over2T^2}\right].
\end{eqnarray}
Here $E_{\bf k}^s=E_{\bf k}+s\delta\mu$ for convenience.

For $s$-wave or $p$-wave pairing, we only need to change the corresponding gamma functions $\Gamma^{\rm d}({\bf k})$ to $\Gamma^{\rm s,p}({\bf k})$ and set $\delta\mu\equiv0$ for the $p$-wave case. 
For $s$-wave pairing where $\Gamma^{\rm s}({\bf k})=1$,   Eq.(\ref{F2}) becomes
\begin{eqnarray}
\partial_{q_i}^2F({\bf q})|_{\bf q=0}&=&\sum_{{\bf k},m}\left\{{\xi_{\bf k}\partial_{k_i}^2\xi_{\bf k}+3(\partial_{k_i}\xi_{\bf k})^2\over\Big[(i\tilde\omega_m)^2-E_{\bf k}^2\Big]^2}+{4(\xi_{\bf k}\partial_{k_i}\xi_{\bf k})^2\over\Big[(i\tilde\omega_m)^2-E_{\bf k}^2\Big]^3}\right\}.
\end{eqnarray}
Using the following identities
\begin{eqnarray}
-{\partial\over\partial\mu}{\xi_{\bf k}\over\Big[(i\tilde\omega_m)^2-E_{\bf k}^2\Big]^2}&=&{1\over\Big[(i\tilde\omega_m)^2-E_{\bf k}^2\Big]^2}+{4\xi_{\bf k}^2\over\Big[(i\tilde\omega_m)^2-E_{\bf k}^2\Big]^3},\\
-\sum_{{\bf k},m}(\partial_{k_i}\xi_{\bf k})^2{\partial\over\partial\mu}{\xi_{\bf k}\over\Big[(i\tilde\omega_m)^2-E_{\bf k}^2\Big]^2}&=&\sum_{{\bf k},m}(\partial_{k_i}\xi_{\bf k})^2{\partial\over\partial\xi_{\bf k}}{\xi_{\bf k}\over\Big[(i\tilde\omega_m)^2-E_{\bf k}^2\Big]^2}={2m\over4\pi}\sum_m\int_0^\infty d\xi_{\bf k}(\partial_{k_i}\xi_{\bf k})^2{\partial\over\partial\xi_{\bf k}}{\xi_{\bf k}\over\Big[(i\tilde\omega_m)^2-E_{\bf k}^2\Big]^2}\nonumber\\
&=&-{2m\over4\pi}\sum_m\int_0^\infty d\xi_{\bf k}\partial_{k_i}^2\xi_{\bf k}{\xi_{\bf k}\over\Big[(i\tilde\omega_m)^2-E_{\bf k}^2\Big]^2}=-\sum_{{\bf k},m}\partial_{k_i}^2\xi_{\bf k}{\xi_{\bf k}\over\Big[(i\tilde\omega_m)^2-E_{\bf k}^2\Big]^2},
\end{eqnarray}
we obtain
\begin{eqnarray}
\partial_{q_i}^2F({\bf q})|_{\bf q=0}=\sum_{{\bf k},m}{2(\partial_{k_i}\xi_{\bf k})^2\over\Big[(i\tilde\omega_m)^2-E_{\bf k}^2\Big]^2}=\sum_{{\bf k}}{{\bf k}^2\over 4m^2E_{\bf k}^2}\left[{\tanh\left({E_{\bf k}\over2T}\right)\over E_{\bf k}}-{{\rm sech}^2\left({E_{\bf k}\over2T}\right)\over2T}\right]
\end{eqnarray}
at $\delta\mu=0$. This is equivalent to the explicit form given in~\cite{Botelho2006}.
\end{widetext}

\end{document}